\documentclass[conference]{IEEEtran}
\IEEEoverridecommandlockouts
\usepackage{cite}
\usepackage{amsmath,amssymb,amsfonts}
\usepackage{algorithmic}
\usepackage{graphicx}
\usepackage{subcaption}
\usepackage{textcomp}
\usepackage{xcolor}
\usepackage[caption=false,font=footnotesize]{subfig}
\usepackage{algorithmic}
\usepackage{algorithm}
\usepackage{amsmath,bm}
\usepackage{multirow}
\usepackage{threeparttable}
\usepackage{caption}
\usepackage{graphicx}
\usepackage{float} 
\usepackage{subcaption}

\def\BibTeX{{\rm B\kern-.05em{\sc i\kern-.025em b}\kern-.08em
		T\kern-.1667em\lower.7ex\hbox{E}\kern-.125emX}}
\begin{document}
	\title{Ground roll suppression based on the synchrosqueezed wavelet transform and local adaptive subtraction filtering}
    \author{Zifei Li, Shaohuan Zu, and Zhengyu Tang%
    \thanks{This research is partly supported by the National Natural Science Foundation of China (Grant No. 42374160), CNPC Innovation Fund (2024DQ02-0139) and High-performance Computing Platform of Chengdu University of Technology.}
    \thanks{Zifei Li, Shaohuan Zu, and Zhengyu Tang are with the State Key Laboratory of Oil and Gas Reservoir Geology and Exploitation, Chengdu University of Technology, Chengdu 610059, China, and also with the College of Geophysics, Chengdu University of Technology, Chengdu 610059, China.}%
    \thanks{Shaohuan Zu and Zhengyu Tang is also with the Key Laboratory of Earth Exploration and Information Technology of Ministry of Education, Chengdu University of Technology, Chengdu 610059, China.}
    \thanks{Corresponding author: Shaohuan Zu (email: zushaohuan19@cdut.edu.cn). }%
    }
	\maketitle
	\begin{abstract}
Ground roll suppression remains a challenge in land seismic data processing, particularly when strong surface wave energy exhibits significant overlap with effective reflection events in the low frequency band. Conventional approaches, such as frequency-wavenumber(F-K) filtering, Radon transform-based methods, and other spatial filtering techniques, often fail to achieve satisfactory results under such conditions due to limited resolution and poor amplitude preservation. To address this problem, a time-frequency domain ground roll suppression method that integrates the synchrosqueezed wavelet transform (SSWT) with local adaptive subtraction filtering is proposed. First, seismic traces are transformed from the time-space domain into the time-frequency domain using SSWT, which provides a highly concentrated time-frequency representation. Based on the distinct distribution characteristics of ground roll noise and effective reflection signals in the time-frequency domain, a ground roll model is extracted. Subsequently, a local adaptive subtraction filter is applied to adjust the amplitude and phase of the extracted ground roll model, and the corrected model is subtracted from the original seismic data to obtain the final suppressed result. Applications to both synthetic and field seismic data demonstrate that the proposed method effectively attenuates ground roll energy while maximally preserving the amplitude and waveform characteristics of reflection signals. Compared with conventional time-frequency-based ground roll suppression approaches, the proposed method exhibits superior amplitude fidelity and improved noise attenuation performance.
	\end{abstract}
	
	\begin{IEEEkeywords}
		seismic data processing; ground roll suppression; time-frequency analysis; synchrosqueezed wavelet transform; adaptive subtraction filtering
	\end{IEEEkeywords}
	\section{Introduction}
        Ground roll is one of the most typical and dominant types of coherent noise encountered in land seismic data. It is characterized by strong energy, low apparent velocity, low-frequency content, and pronounced dispersion, which severely contaminate the desired reflection signals in both shallow near-offset zones and deeper far-offset regions. Consequently, suppressing ground roll while preventing the loss of desired signals across different frequency bands has long been a pivotal research topic in high-resolution and high-signal-to-noise-ratio (SNR) seismic exploration \cite{r19,r20}.

        Based on the kinematic and dynamic characteristics of ground roll, traditional suppression methods primarily include high-pass filtering, $f-k$ filtering, singular value decomposition (SVD) filtering, the Radon transform, the Karhunen-Loève transform, and the curvelet transform. Although these methods can achieve acceptable results under specific conditions, they generally suffer from limited resolution, insufficient amplitude preservation, and incomplete separation between the ground roll and the desired reflections. In particular, when the ground-roll energy significantly overlaps with reflection signals in the low-frequency band, conventional methods often struggle to optimally balance noise attenuation with signal preservation \cite{r1,r2,r21,r3,r4,r5}.

        Advancements in time-frequency (TF) analysis techniques have facilitated the suppression of ground roll by exploiting the distinct TF distributions of the coherent noise and the desired signals, yielding promising results. Deighan et al. \cite{r6} applied the 1-D discrete wavelet transform to ground-roll suppression, demonstrating superior performance over conventional band-pass filtering. Askari and Siahkoohi \cite{r7} proposed an attenuation workflow combining the S-transform with $x-f-k$ filtering, which significantly improved noise suppression while preserving the primary reflections. Furthermore, an adaptive approach based on the continuous wavelet transform was developed by Chen et al. \cite{r8} to exploit the offset-variant characteristics of ground roll. However, the efficacy of wavelet-based methods is heavily contingent upon the empirical selection of the mother wavelet, which can lead to inconsistent separation outcomes. To address this limitation, Li et al. \cite{r9} introduced a generalized S-transform-based technique, attenuating ground-roll energy via localized TF filtering. Additionally, the Hilbert-Huang transform \cite{r10} and the Gabor transform \cite{r11} have been successfully employed to segment seismic traces and isolate ground roll based on amplitude discrepancies within their respective transformed domains.

        The efficacy of most time-frequency-based (TF-based) ground-roll suppression methods heavily relies on the design of localized filtering windows, wherein TF resolution plays a critical role. However, conventional TF analysis tools, such as the continuous wavelet transform and the generalized S-transform, are intrinsically restricted by the Heisenberg uncertainty principle, which inevitably limits their achievable TF resolution. To overcome this limitation, the synchrosqueezing algorithm was introduced into the wavelet transform by Daubechies et al. \cite{r12}, significantly enhancing the TF concentration. Subsequently, the synchrosqueezed wavelet transform was successfully applied to ground-roll attenuation \cite{r13}. To further optimize the separation of ground roll and desired reflections, recent studies have exploited the frequency-shifting property of the Fourier transform to project low-frequency energy into higher frequency bands within the synchrosqueezed wavelet domain \cite{r14}, or extended the SST to the spatial domain to leverage joint spatio-temporal synchrosqueezing \cite{r15}. Despite these advancements, since most TF methods primarily operate on a trace-by-trace basis, they remain computationally intensive during the empirical window-design phase and inherently neglect the vital spatial correlations among adjacent seismic traces.

        In this study, a novel attenuation framework integrating the synchrosqueezed wavelet transform (SSWT) and an adaptive subtraction scheme is proposed. Specifically, the ground-roll energy is initially suppressed within the time-frequency domain via a tailored thresholding strategy, which optimally preserves the desired reflections. Subsequently, a localized adaptive subtraction filter is employed to meticulously calibrate the amplitude and phase of the estimated ground-roll model, thereby facilitating amplitude-preserving noise attenuation. Extensive evaluations on both synthetic and field datasets, benchmarked against conventional methods, demonstrate that the proposed approach yields significant improvements in the signal-to-noise ratio (SNR) while maintaining strict amplitude fidelity.

\section{Methods}

\subsection{Synchrosqueezed wavelet transform}
            Conventional time-frequency (TF) analysis methods in signal processing primarily include the short-time Fourier transform (STFT), the continuous wavelet transform (CWT), the S-transform, and the Wigner-Ville distribution (WVD) \cite{r22,r23,r24,r25}. However, the achievable TF resolution of these linear transforms is intrinsically restricted by the Heisenberg uncertainty principle, whereas quadratic representations, such as WVD-type methods, inevitably suffer from severe cross-term interference \cite{r26}.

            To address these limitations, the synchrosqueezed wavelet transform (SSWT), a TF reassignment technique based on the CWT, was introduced by Daubechies et al. \cite{r12}. Conceptually analogous to empirical mode decomposition (EMD), the SSWT provides a fully data-driven representation while benefiting from a rigorous mathematical foundation \cite{r16}. The core mechanism of the SSWT involves reallocating the continuous wavelet coefficients along the frequency axis based on the computed instantaneous frequency. This reassignment yields a highly concentrated TF representation with significantly enhanced resolution. Compared with the conventional CWT, the SSWT provides a sharper characterization of instantaneous frequency variations for nonstationary signals, rendering it exceptionally well-suited for the analysis of multicomponent time-varying signals, such as seismic records.

            For a given 1-D signal $s(t)$, the continuous wavelet transform (CWT) is defined as:
            \begin{equation}
                W_s(a,b)=\int_{-\infty}^{+\infty} s(t) a^{-1/2} \overline{\psi\left(\frac{t-b}{a}\right)} \,\mathrm{d}t,
                \label{eq:cwt_def}
            \end{equation}
            where $\psi(t)$ denotes the mother wavelet, $\overline{\psi(t)}$ represents its complex conjugate, $a$ defines the scale parameter, and $b$ indicates the translation parameter. 

            Substituting a purely harmonic signal, $s(t)=A \cos (\omega t)$, into equation~\ref{eq:cwt_def} and applying Plancherel's theorem yields the following frequency-domain representation:
            \begin{equation}
            \begin{aligned}
                W_s(a, b) & =\frac{1}{2 \pi} \int_{-\infty}^{+\infty} \widehat{s}(\xi) a^{-1 / 2} \overline{\widehat{\psi}(a \xi)} e^{i b \xi} \,\mathrm{d}\xi \\
                & =\frac{A}{2} \int_{-\infty}^{+\infty} [\delta(\xi-\omega)+\delta(\xi+\omega)] a^{-1 / 2} \overline{\widehat{\psi}(a \xi)} e^{i b \xi} \,\mathrm{d}\xi \\
                & =\frac{A}{2} a^{-1 / 2} \overline{\widehat{\psi}(a \omega)} e^{i b \omega},
            \end{aligned}
            \label{eq:cwt_harmonic}
            \end{equation}
            where $\xi$ denotes the angular frequency, while $\widehat{s}$ and $\widehat{\psi}$ represent the Fourier transforms of $s(t)$ and $\psi(t)$, respectively. Note that the negative frequency component strictly vanishes in the final derivation step under the assumption that an analytic mother wavelet (i.e., $\widehat{\psi}(\xi) = 0$ for $\xi < 0$) is employed.

            Assuming that the frequency spectrum of the mother wavelet, $\widehat{\psi}(\xi)$, is highly concentrated around a central frequency $\xi=\omega_0$, the continuous wavelet coefficients $W_{s}(a,b)$ are principally mapped to the scale $a=\omega_0/\omega$. However, owing to the inherent energy leakage in the frequency domain, the resultant coefficient bands within the time-frequency (TF) representation inevitably suffer from spectral broadening and blurred boundaries, thereby severely compromising the achievable TF resolution. To rigorously address this limitation, Daubechies et al. \cite{r12} demonstrated that the instantaneous frequency can be explicitly estimated by computing the partial derivative of $W_{s}(a,b)$ with respect to the translation parameter $b$ (i.e., time), which is mathematically formulated as:
            \begin{equation}
                \omega_s(a,b)=
                \begin{cases}
                    -\mathrm{i}\dfrac{\partial_b W_s(a,b)}{W_s(a,b)}, & W_s(a,b)\neq 0, \\[6pt]
                    \infty, & W_s(a,b)=0.
                \end{cases}
                \label{eq:inst_freq}
            \end{equation}

            Because the continuous wavelet transform initially maps the 1-D signal into the time-scale domain $(b,a)$, the resultant wavelet coefficients must be reassigned to the TF domain $(b,\omega_{s}(a,b))$ to execute the synchrosqueezing operation \cite{r27}. Through a rigorous reassignment algorithm, the wavelet coefficients within a localized neighborhood of an arbitrary center frequency $\omega_{l}$, defined as $\left[\omega_{l}-\frac{1}{2}(\omega_{l}-\omega_{l-1}), \omega_{l}+\frac{1}{2}(\omega_{l}-\omega_{l-1})\right]$, are systematically reallocated onto the center frequency. This procedure yields the synchrosqueezed wavelet coefficients $T_{s}(\omega_{l},b)$, thereby achieving a significantly enhanced frequency resolution. Given that both the scale $a$ and the angular frequency $\omega$ are discretely sampled, let $\Delta\omega=\omega_{l}-\omega_{l-1}$ and $(\Delta a)_{k}=a_{k}-a_{k-1}$, where $l$ and $k$ denote the discrete frequency and scale indices, respectively. The explicit mathematical expression for the discrete synchrosqueezed wavelet coefficients $T_s(\omega_{l},b)$ is formulated as:
            \begin{equation}
                T_s(\omega_l,b) = (\Delta\omega)^{-1} \sum_{a_k: |\omega_s(a_k,b)-\omega_l| \leq \Delta\omega/2} W_s(a_k,b) a_k^{-3/2} (\Delta a)_k.
                \label{eq:sswt_discrete}
            \end{equation}

\begin{figure}
                \centering
                \begin{subfigure}{1\linewidth}
                    \centering
                    \includegraphics[width=1\linewidth]{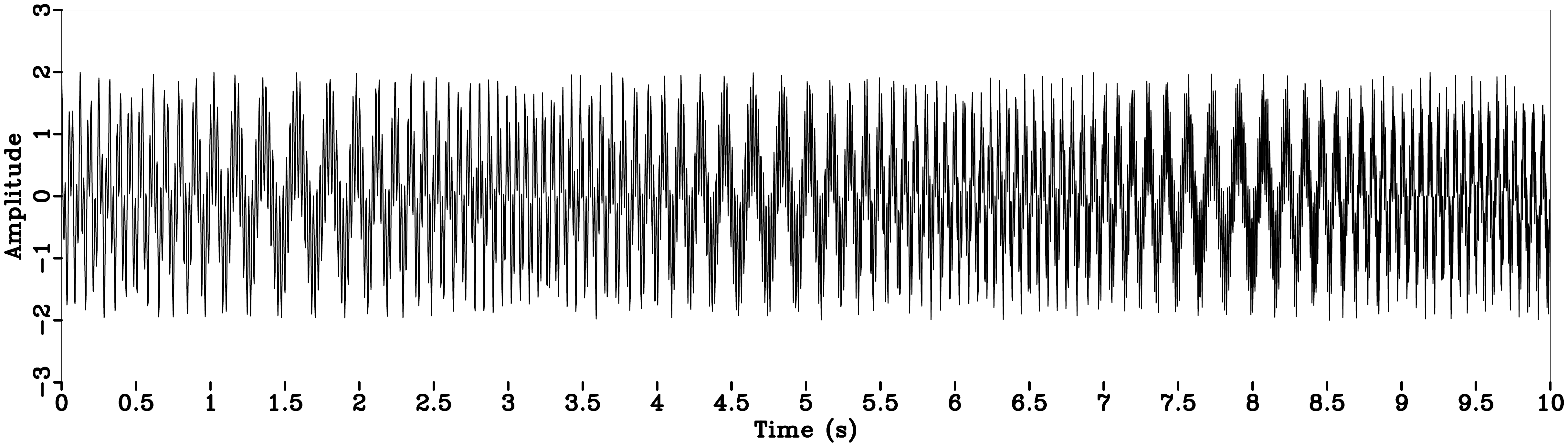}
                    \caption{Synthetic signal}
                    \label{fig:synthetic_signal}
                \end{subfigure}
                \par\medskip
                \begin{subfigure}{0.49\linewidth}
                    \centering
                    \includegraphics[width=1\linewidth]{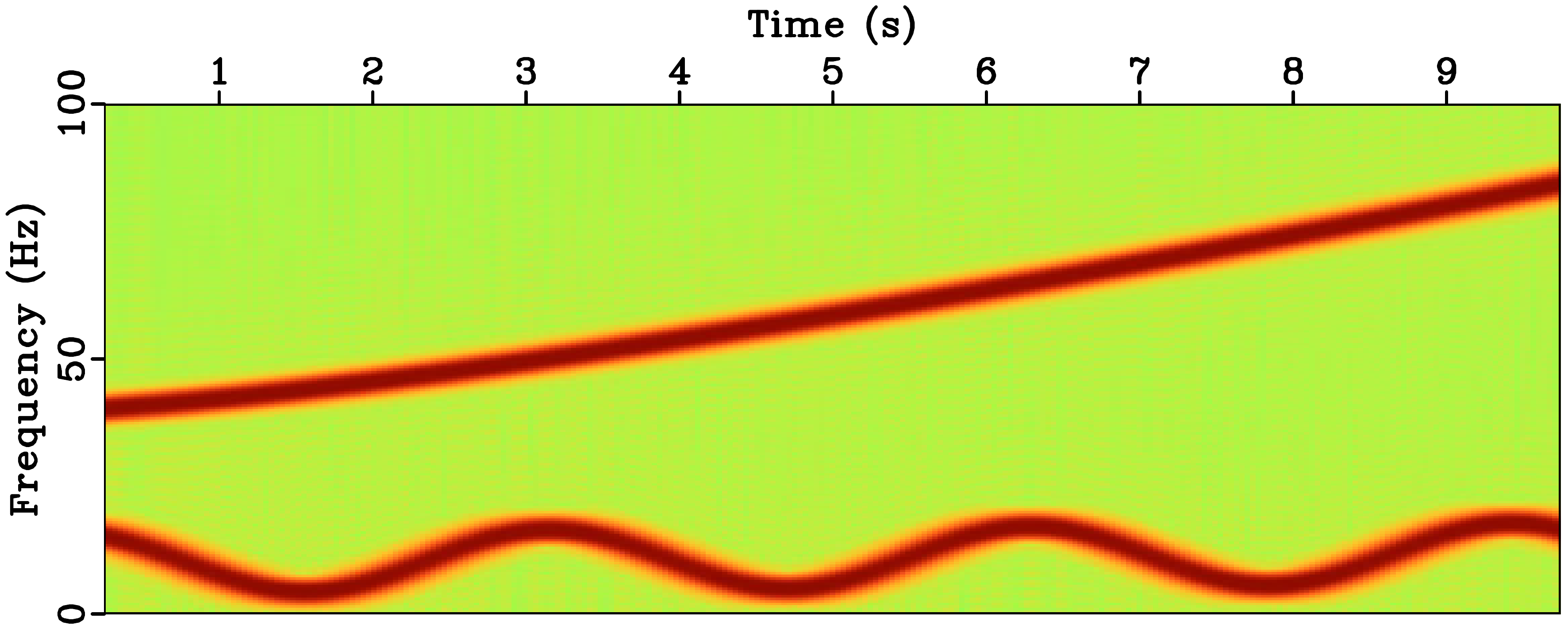}
                    \caption{STFT}
                    \label{fig:tf_stft}
                \end{subfigure}
                \begin{subfigure}{0.49\linewidth}
                    \centering
                    \includegraphics[width=1\linewidth]{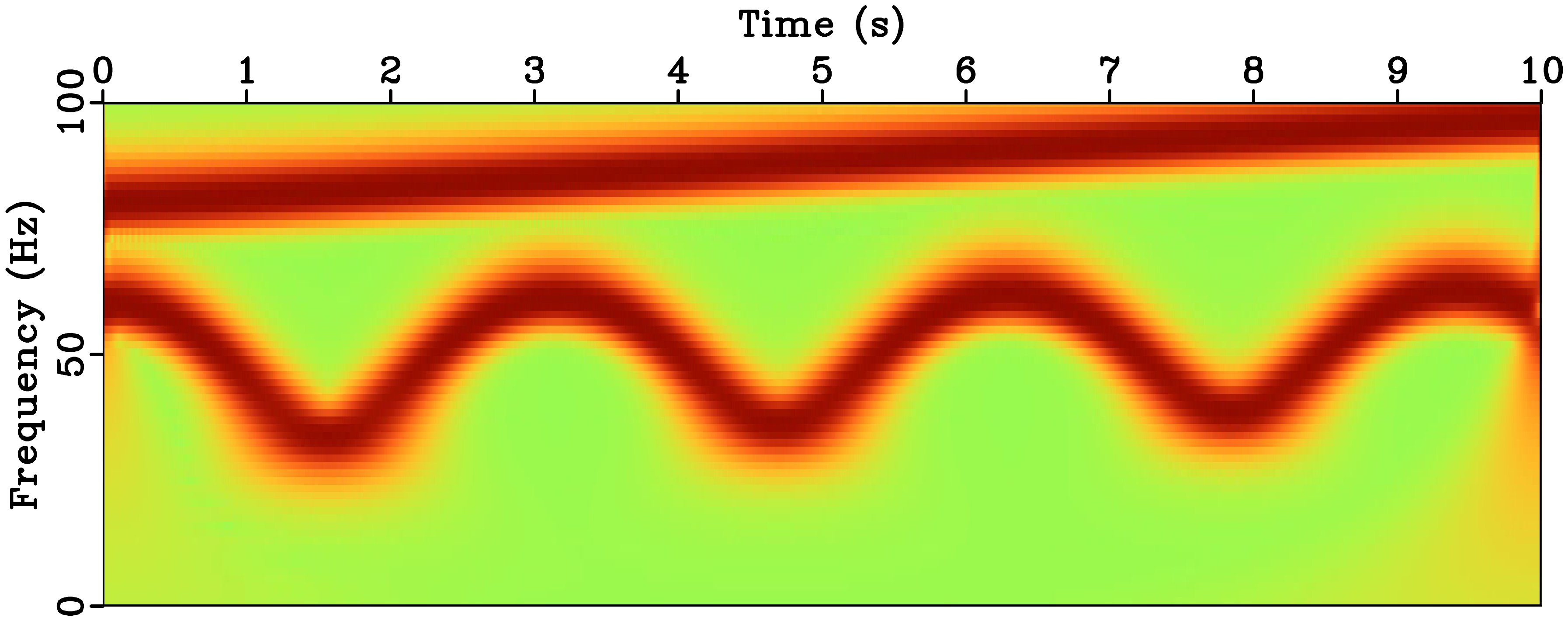}
                    \caption{CWT}
                    \label{fig:tf_cwt}
                \end{subfigure}
                \par\medskip
                \begin{subfigure}{0.49\linewidth}
                    \centering
                    \includegraphics[width=1\linewidth]{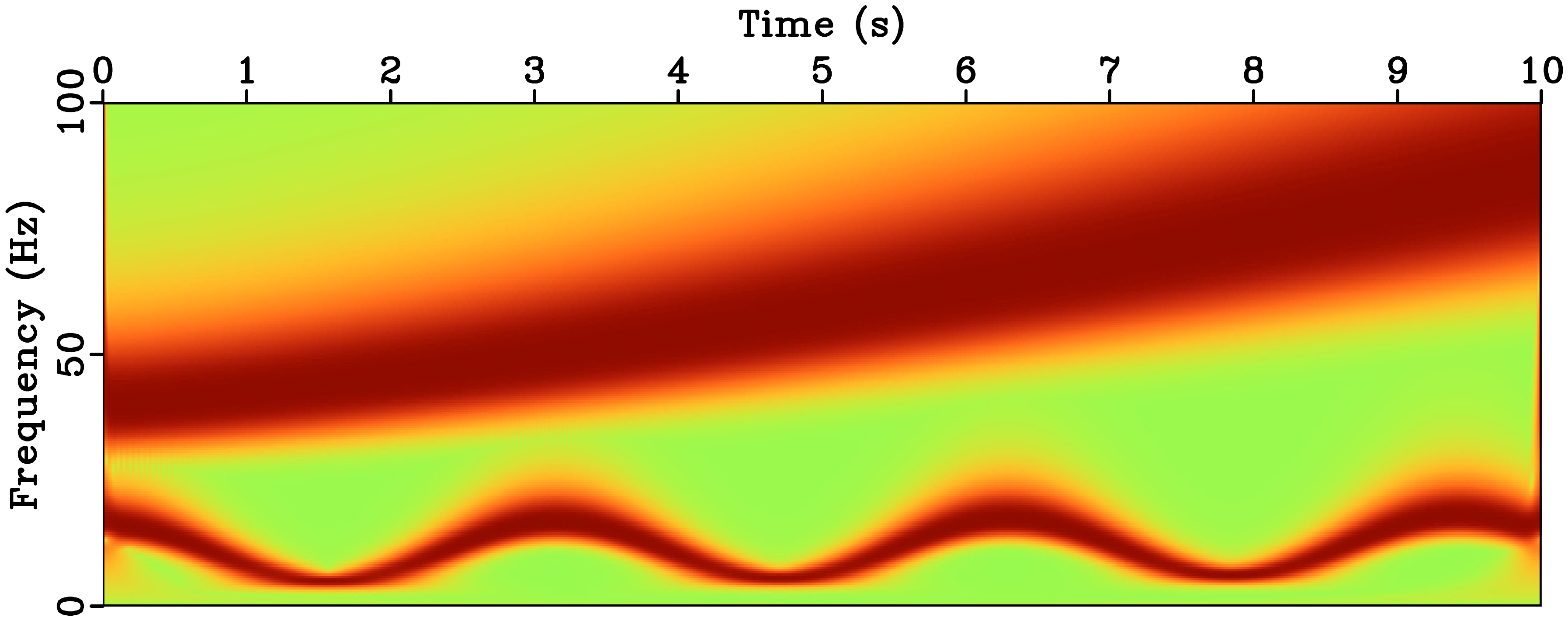}
                    \caption{ST}
                    \label{fig:tf_st}
                \end{subfigure}
                \begin{subfigure}{0.49\linewidth}
                    \centering
                    \includegraphics[width=1\linewidth]{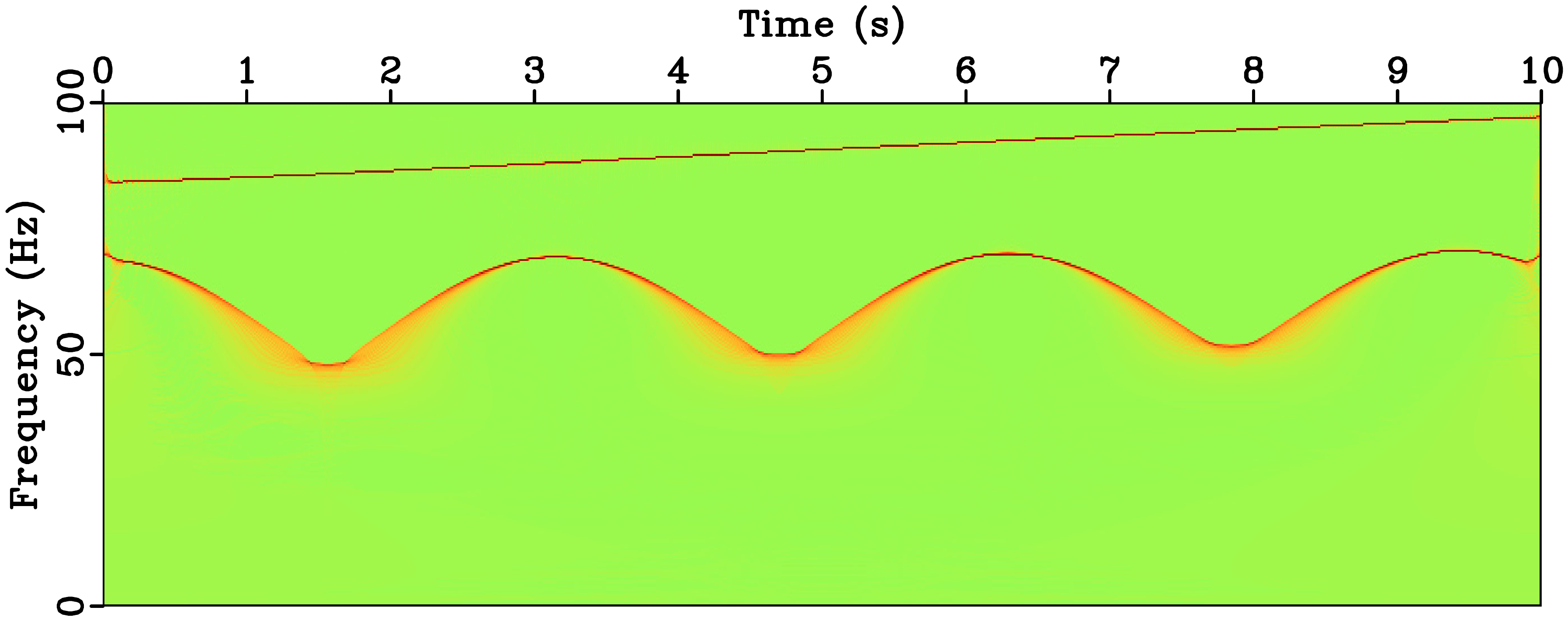}
                    \caption{SSWT}
                    \label{fig:tf_sswt}
                \end{subfigure}
                \caption{Synthetic signal and time-frequency spectra obtained using different methods.}
                \label{fig:method}
            \end{figure}

            To explicitly demonstrate the superiority of the SSWT in TF analysis, a synthetic composite signal comprising two distinct components is constructed as $s(t)=s_{1}(t)+s_{2}(t)$, given by:
            \begin{equation}
            \begin{cases}
                s_1(t)=\cos\left[2\pi(0.1t^2+3\sin(2t)+10t)\right], \\
                s_2(t)=\cos\left[2\pi((40+t^{1.3})t)\right],
            \end{cases}
            \label{eq:synthetic_signal}
            \end{equation}
            where $s_1(t)$ represents a low-frequency nonstationary component characterized by slow frequency and sinusoidal modulations, whereas $s_2(t)$ serves as a high-frequency component with a monotonically increasing frequency over time. The synthetic signal and its corresponding TF representations, computed via various analytical methods, are illustrated in Figure~\ref{fig:method}. As observed, the STFT is intrinsically restricted by the predetermined window function, resulting in a compromised TF resolution across both the low- and high-frequency bands. Although the CWT and S-transform provide superior low-frequency resolution compared to the STFT, both inevitably suffer from severe energy leakage at higher frequencies. In stark contrast, the SSWT exhibits a highly concentrated energy distribution and a substantially enhanced TF resolution.

            It has been proven that the forward and inverse transforms of SSWT are lossless, enabling signal reconstruction\cite{r16}. The inverse transform is given in:
            \begin{equation}
            \begin{aligned}
            s(b) \approx {} & \Re\!\left[
            C_\psi^{-1}
            \sum_k
            W_s\!\left(a_k,b\right)
            a_k^{-3/2}
            (\Delta a)_k
            \right] \\
            = {} & \Re\!\left[
            C_\psi^{-1}
            \sum_l
            T_s\!\left(\omega_l,b\right)
            (\Delta \omega)
            \right].
            \end{aligned}
            \end{equation}
            where $C_{\psi}^{-1}=\int_{0}^{+\infty}\overline{\widehat{\psi}(\xi})\frac{\mathrm{d}\xi}{\xi}$. Therefore, SSWT itself does not introduce signal distortion during ground roll suppression.

        \subsection{Local adaptive subtraction filtering}
            Ground roll suppression can be regarded as a model-subtraction problem. Let the observed seismic record be expressed as:
            \begin{equation}
                s(t, x)=z(t, x)-v(t, x)
            \end{equation}
            where $v(t,x)$ denotes the effective signal, and $s(t,x)$ represents the ground roll model. If there exists some discrepancy between the ground roll model and the actual ground roll, direct subtraction often results in residual ground roll or damage to the effective signal. To address this issue, adaptive subtraction filtering employs the least‑squares criterion to optimize the filter operator, adjusting the amplitude and phase of the model locally so that it better matches the original observed data, thereby achieving accurate matching and suppression of ground roll energy \cite{r17}.

            Centered at each time sample $i(i=1,...,M)$, a local short time window containing To achieve spatiotemporal adaptivity, a local window $L$ samples is constructed. A linear convolution relationship is assumed between the ground-roll model and the observed data within the window:
            \begin{equation}
                 z_{i}=S_{i} f_{i}
            \end{equation}
            where $z_{i}$ is the observed data vector inside the window, $S_{i}$ is the local convolution matrix constructed from the ground-roll model $s_{i}$, and $f_{i}$ is the vector of local filter coefficients.

            To solve for the local filter $f_{i}$, a regularized least‑squares estimation is employed, with the objective function given by:
            \begin{equation}
            J(f_i) = \|S_i f_i - z_i\|_2^2 + \lambda \|f_i\|_2^2
            \end{equation}
            where $\lambda>0$ is the regularization parameter used to suppress overfitting and enhance the stability of the solution. The analytical solution for $f_{i}$ is:
            \begin{equation}
            f_i = (s_i^T S_i + \lambda I)^{-1} s_i^T z_i
            \end{equation}
            Using the ground-roll model segment $s_{i}$ corresponding to the center of the window, the predicted ground roll at the current position is:
            \begin{equation}
            \hat{s}_i = f_i^{T} S_i
            \end{equation}
            The final ground-roll‑suppressed seismic record is expressed as:
            \begin{equation}
            \nu_{clean_i} = z_i - \hat{s}_i
            \end{equation}

            By performing the above procedure for each time sample, the complete suppression result is obtained, achieving local adaptive ground-roll suppression.
    \subsection{Ground roll suppression using SSWT and local adaptive subtraction}
            The effectiveness of adaptive subtraction filtering largely depends on the accuracy of the ground roll model. If the model contains reflection energy or lacks complete ground roll components, suppression performance will deteriorate. Therefore, this study extracts the ground roll model based on the distinct energy distribution characteristics of ground roll and reflection signals in the synchrosqueezed wavelet domain and subsequently refines the model using two-dimensional local adaptive subtraction filtering.
            The proposed workflow consists of the following steps:
            
            (1)The original seismic data are separated into high-frequency and low-frequency components using one-dimensional filtering. The high-frequency component is preserved without processing, while the low-frequency component, which contains both ground roll and effective signals, is further processed.
            
            (2)Each low-frequency seismic trace is transformed into the time-frequency domain using SSWT. In this domain, ground roll energy exhibits banded distributions along the time axis, whereas reflection energy is mainly distributed along the frequency axis. A thresholding strategy is applied to retain ground-roll-dominated regions while setting other regions to zero, and the inverse transform is performed to obtain single-trace ground roll models.
            
            (3)The above procedure is repeated for all traces to construct a two-dimensional ground roll model. 
            
            (4)The ground roll model and the low-frequency data are input into the local adaptive subtraction filter to obtain corrected ground roll noise and the suppressed low-frequency effective signals. 
            
            (5)Finally, the suppressed low-frequency component is combined with the preserved high-frequency component to produce the final ground-roll-attenuated seismic data. 
    \section{Numerical modeling tests}
\begin{figure}
            \centering
            \includegraphics[width=0.5\linewidth]{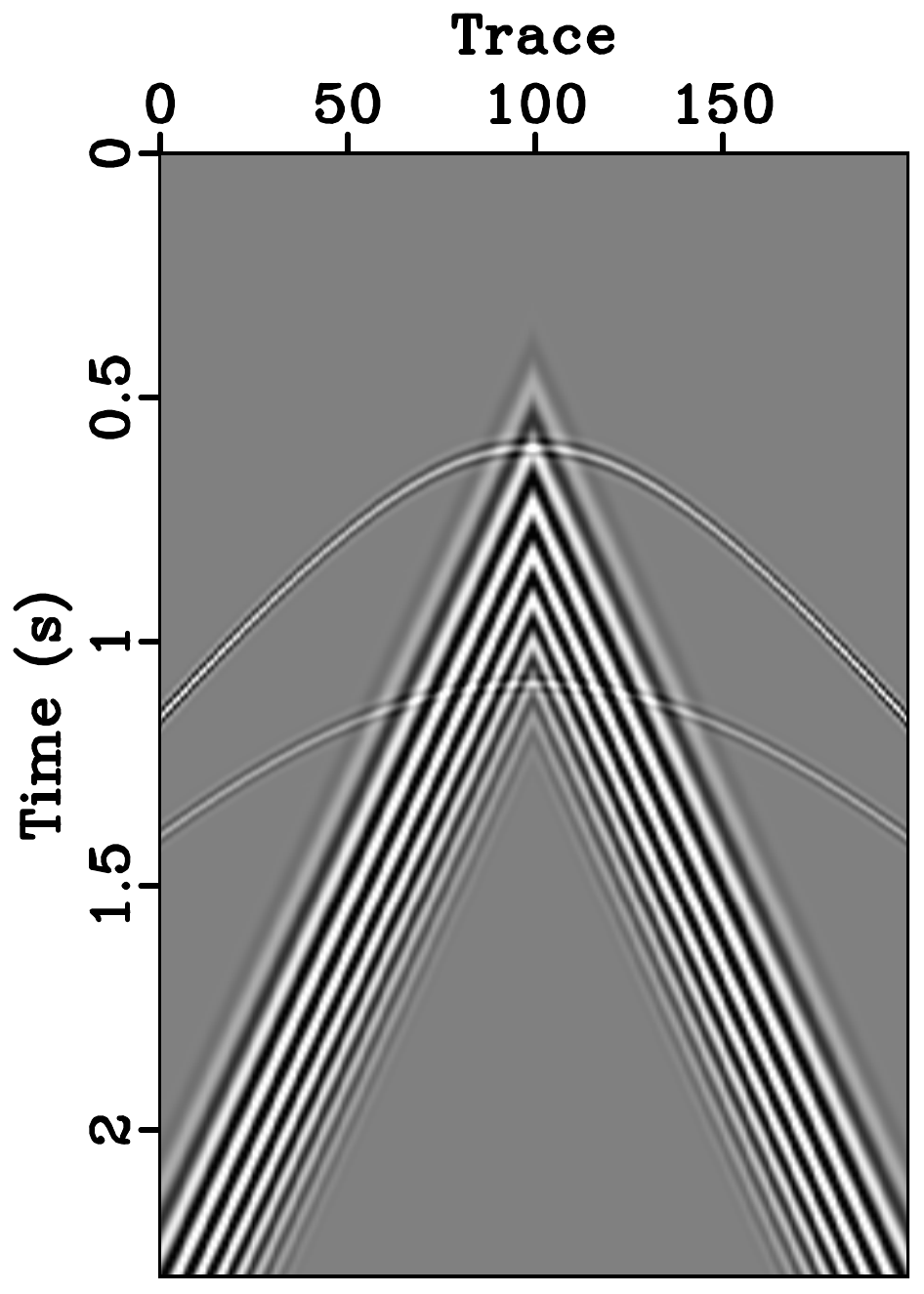}
            \caption{Synthetic seismic data contaminated by ground roll.}
            \label{fig:data_noise}
        \end{figure}

        To rigorously evaluate the efficacy of the proposed method, numerical modeling experiments are initially conducted. Figure~\ref{fig:data_noise} illustrates a single-shot seismic record synthesized via forward modeling using the TesseralPro software, based on a horizontally layered, homogeneous, and isotropic velocity model. The simulated acquisition geometry employs a symmetric shot configuration utilizing a 35-Hz Ricker wavelet as the source signature. The resultant shot gather comprises 200 traces with a spatial interval of 10 m, a temporal sampling rate of 4 ms, and a total recording duration of 2.3 s. The dominant frequency of the generated ground roll is approximately 15 Hz. Furthermore, given that the high-amplitude direct arrivals could severely compromise the subsequent attenuation procedure, they are strictly muted prior to further processing.
        
\begin{figure}
            \centering
            \includegraphics[width=1\linewidth]{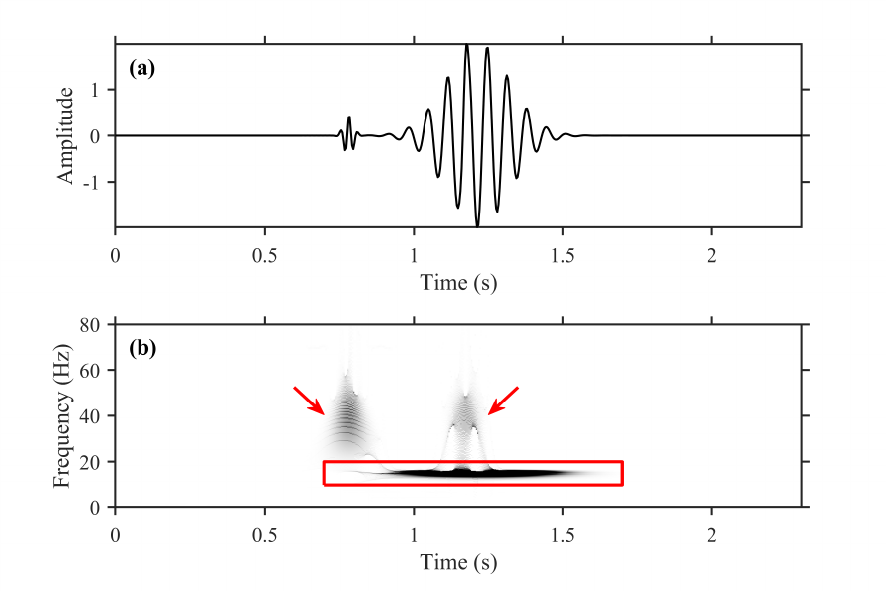}
            \caption{The 50th trace and its time-frequency representation obtained using SSWT:
            (a) seismic trace; (b) SSWT time-frequency spectrum.}
            \label{fig:placeholder}
        \end{figure}

\begin{figure}
            \begin{subfigure}{0.31\linewidth}
                  \centering
                  \includegraphics[width=1\linewidth]{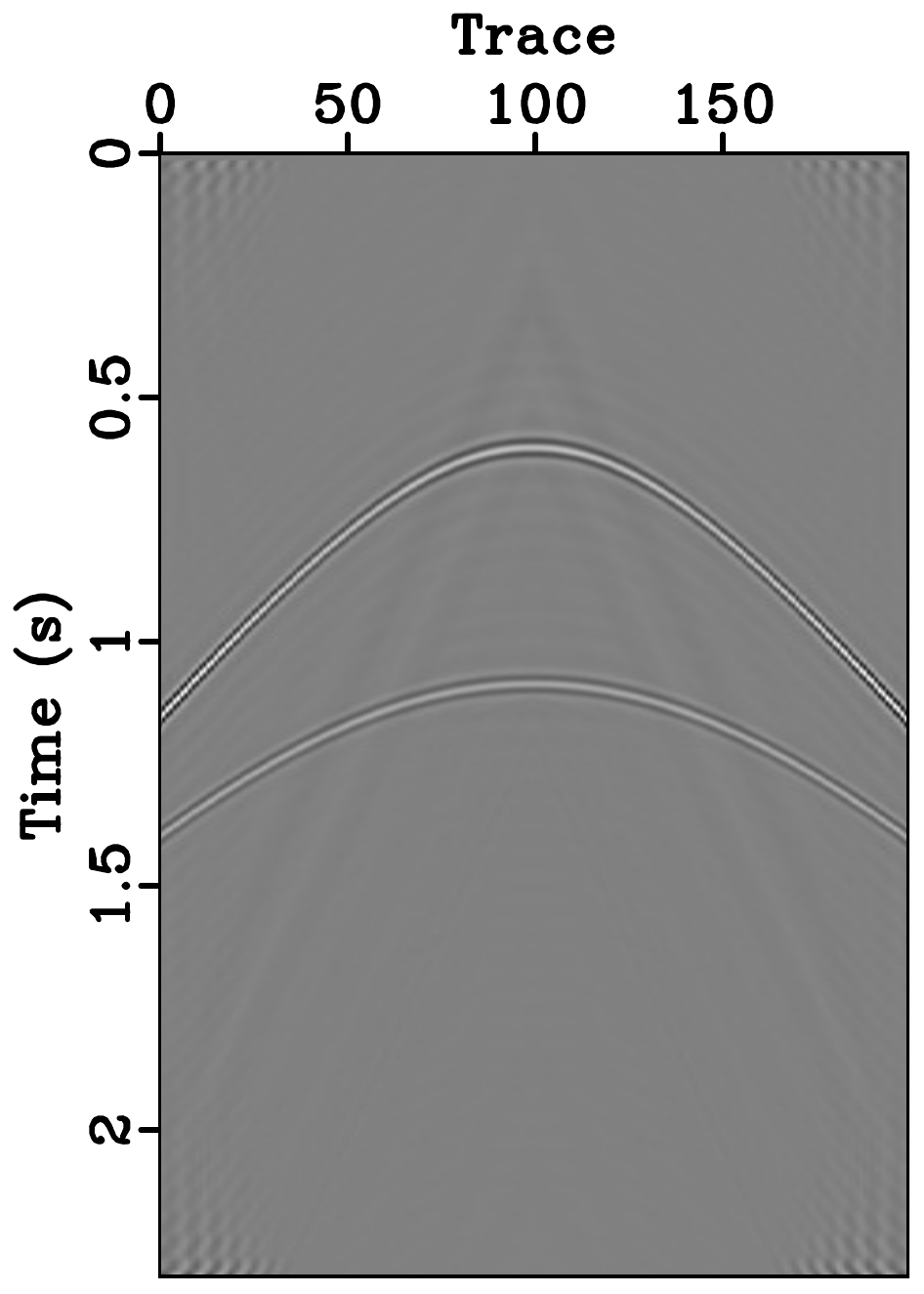}
                  \caption{}
                  \label{fig:data_st}
              \end{subfigure}
              \begin{subfigure}{0.31\linewidth}
                  \centering
                  \includegraphics[width=1\linewidth]{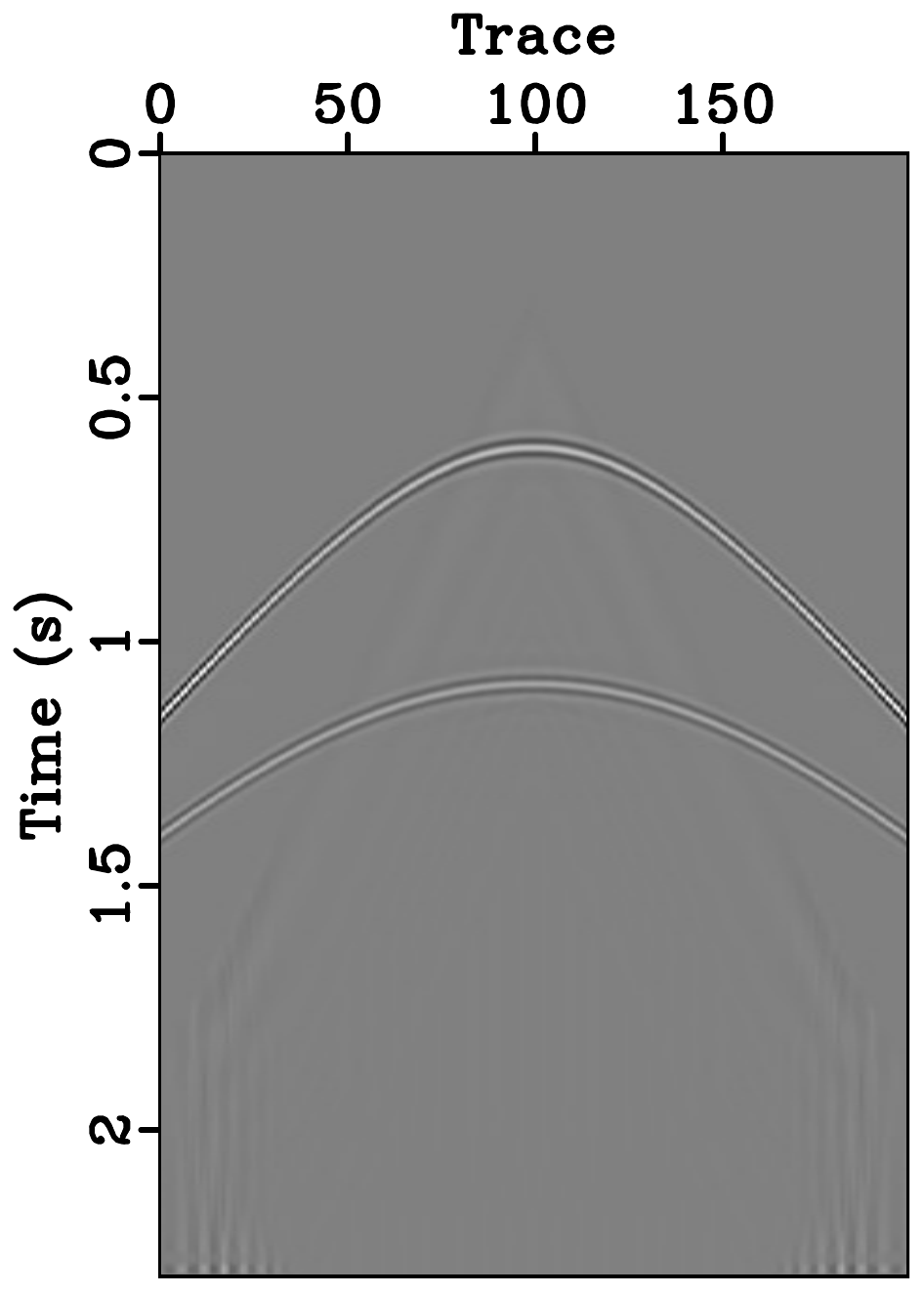}
                  \caption{}
                  \label{fig:data_swt}
              \end{subfigure}
              \begin{subfigure}{0.31\linewidth}
                  \centering
                  \includegraphics[width=1\linewidth]{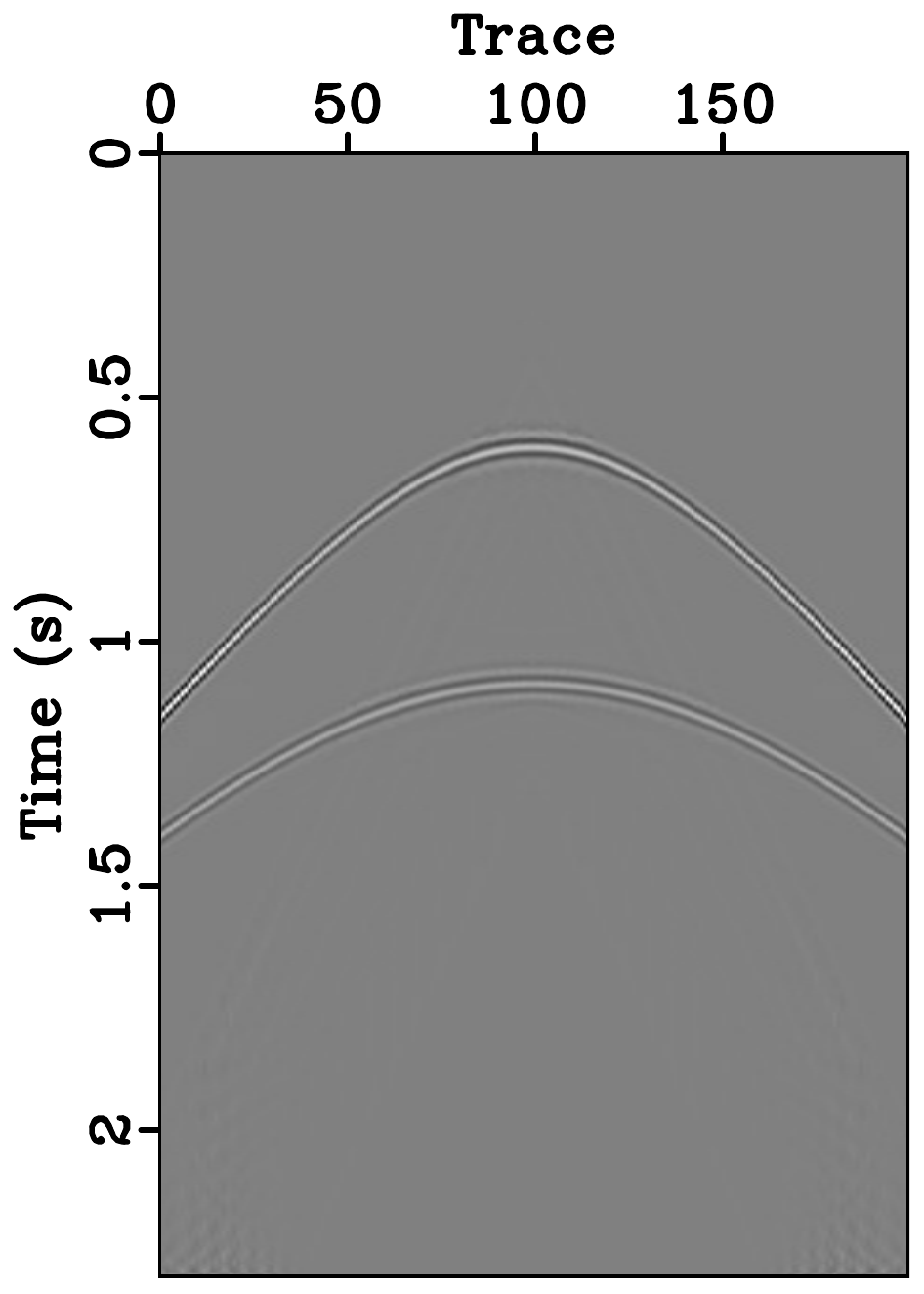}
                  \caption{}
                  \label{fig:data_adapt}
              \end{subfigure}
              \par\medskip
            \begin{subfigure}{0.31\linewidth}
                  \centering
                  \includegraphics[width=1\linewidth]{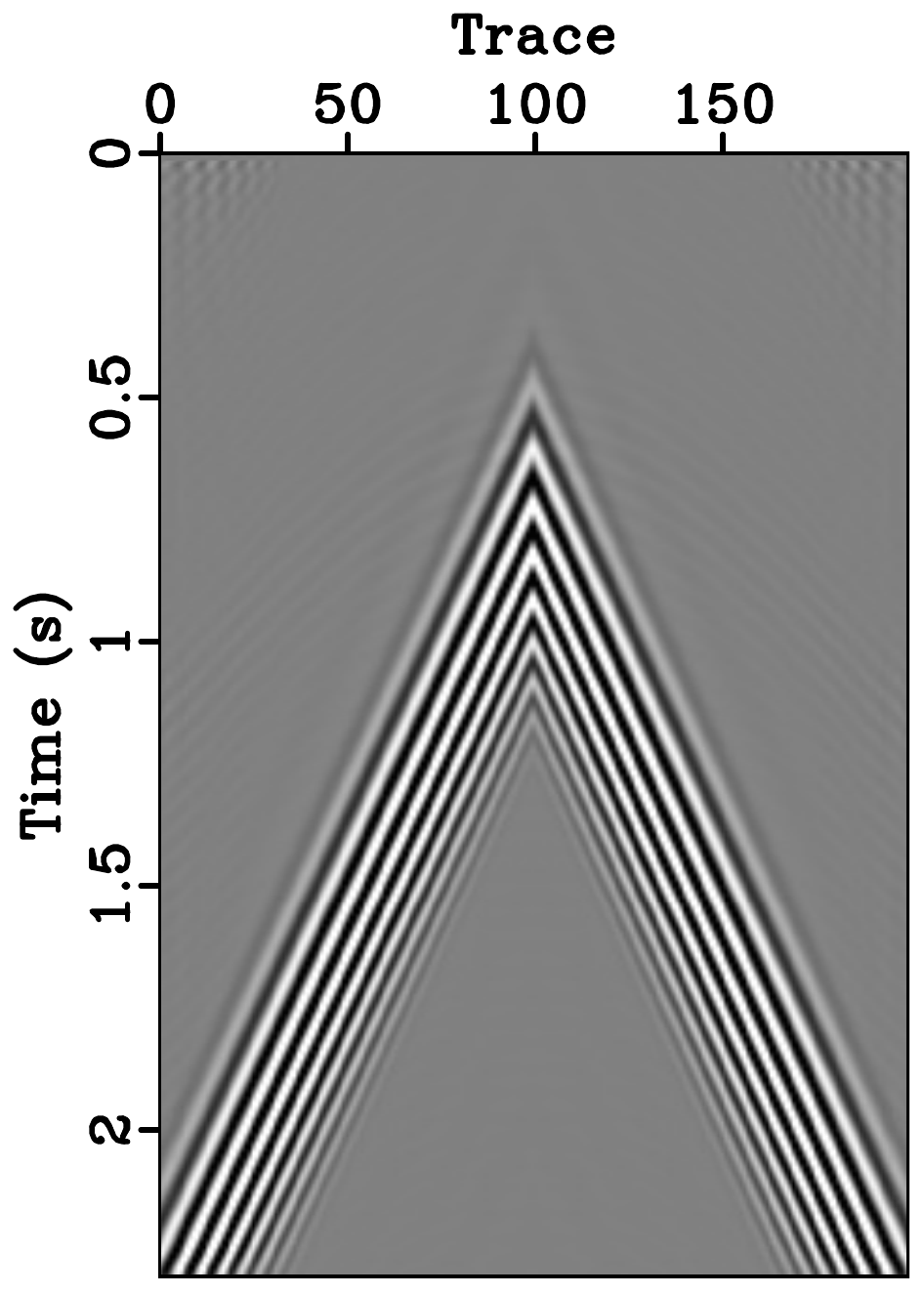}
                  \caption{}
                  \label{fig:data_st_ground}
              \end{subfigure}
              \begin{subfigure}{0.31\linewidth}
                  \centering
                  \includegraphics[width=1\linewidth]{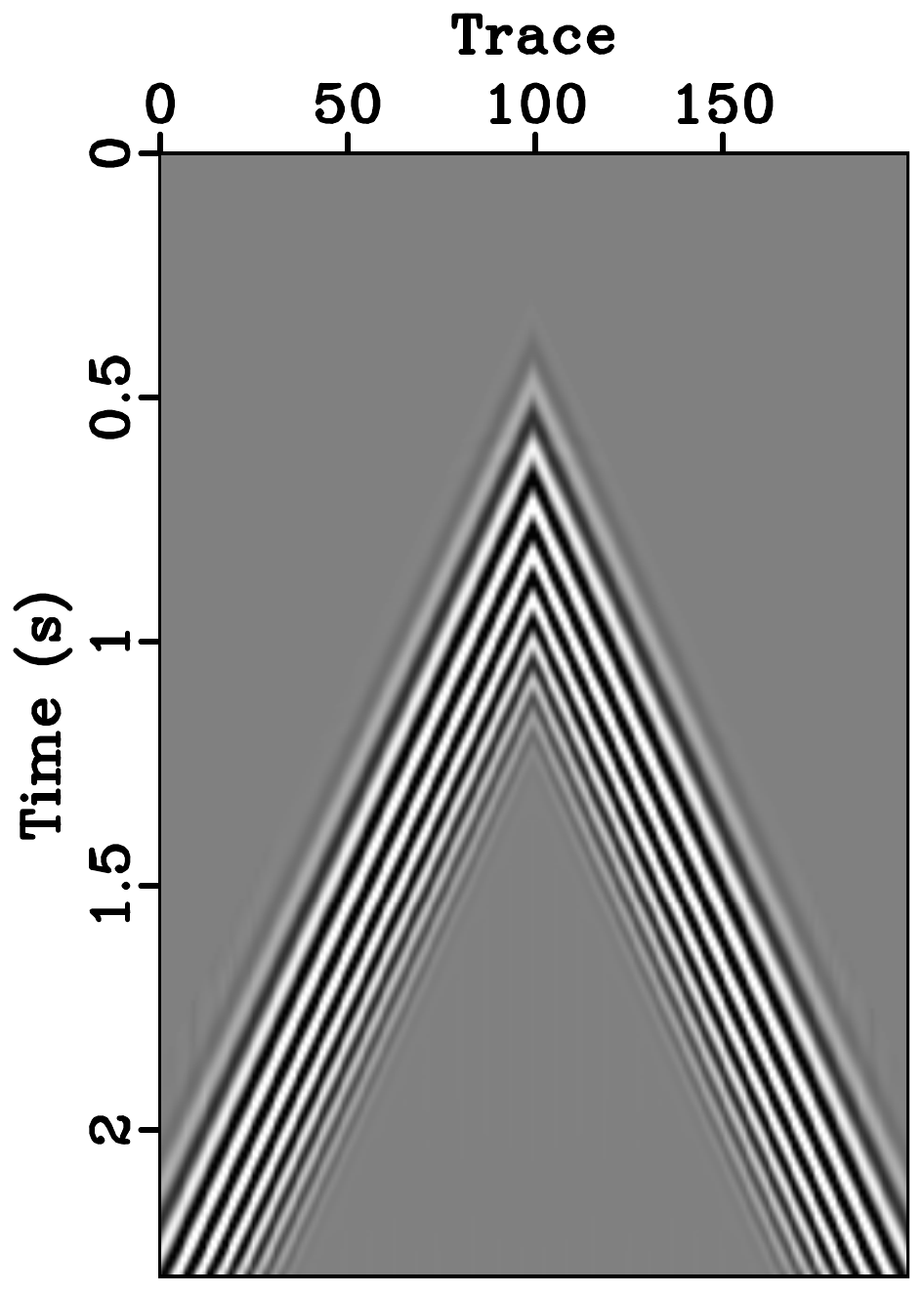}
                  \caption{}
                  \label{fig:data_swt_ground}
              \end{subfigure}
              \begin{subfigure}{0.31\linewidth}
                  \centering
                  \includegraphics[width=1\linewidth]{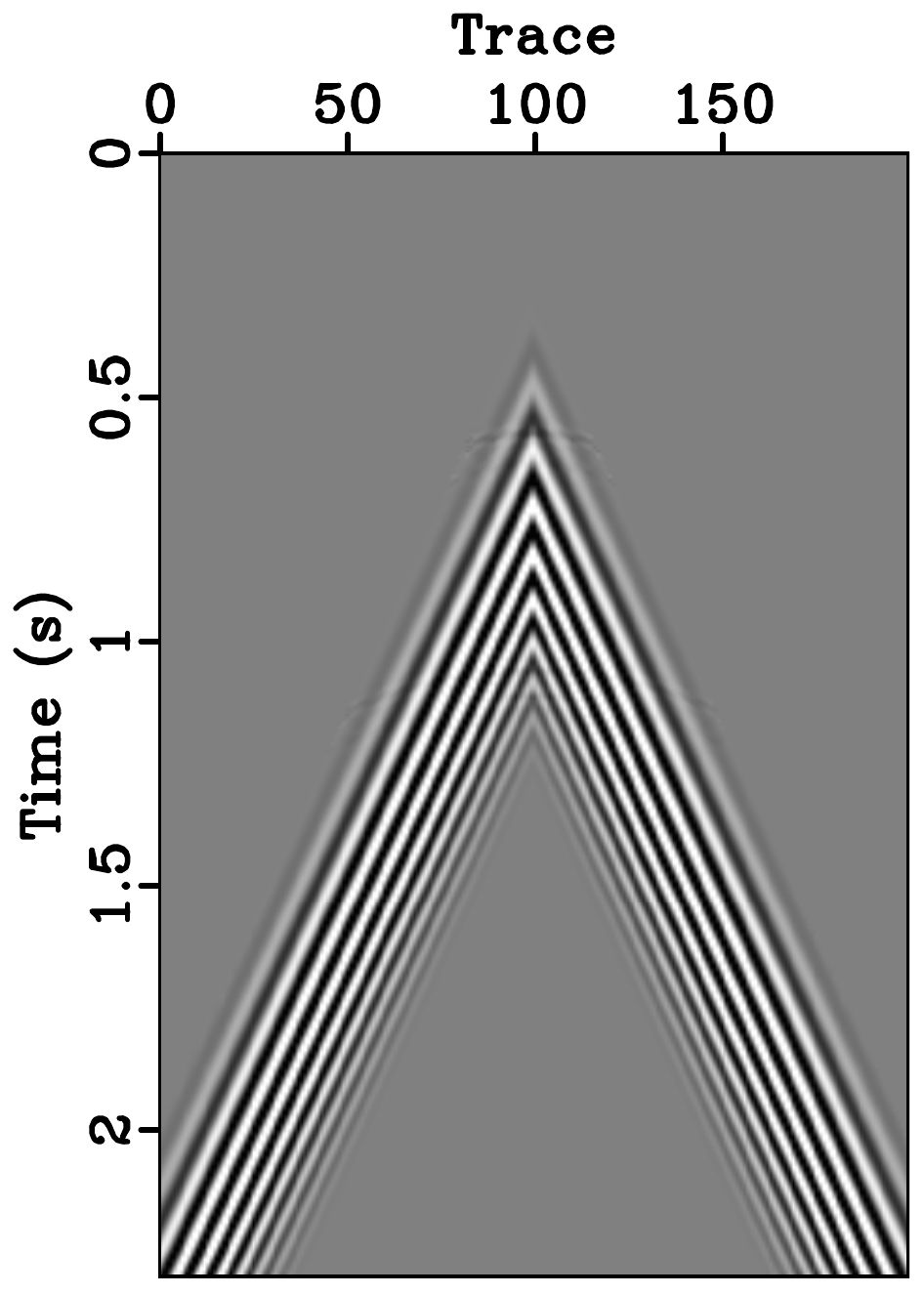}
                  \caption{}
                  \label{fig:data_adapt_ground}
              \end{subfigure}
            \caption{Ground roll suppression results and extracted ground roll components obtained using different methods:(a), (b) S-transform-based method;(c), (d) SSWT-based method;(e), (f) proposed method.}
            \label{fig:Fig_simple}
        \end{figure}

\begin{figure}
            \centering
            \includegraphics[width=1\linewidth]{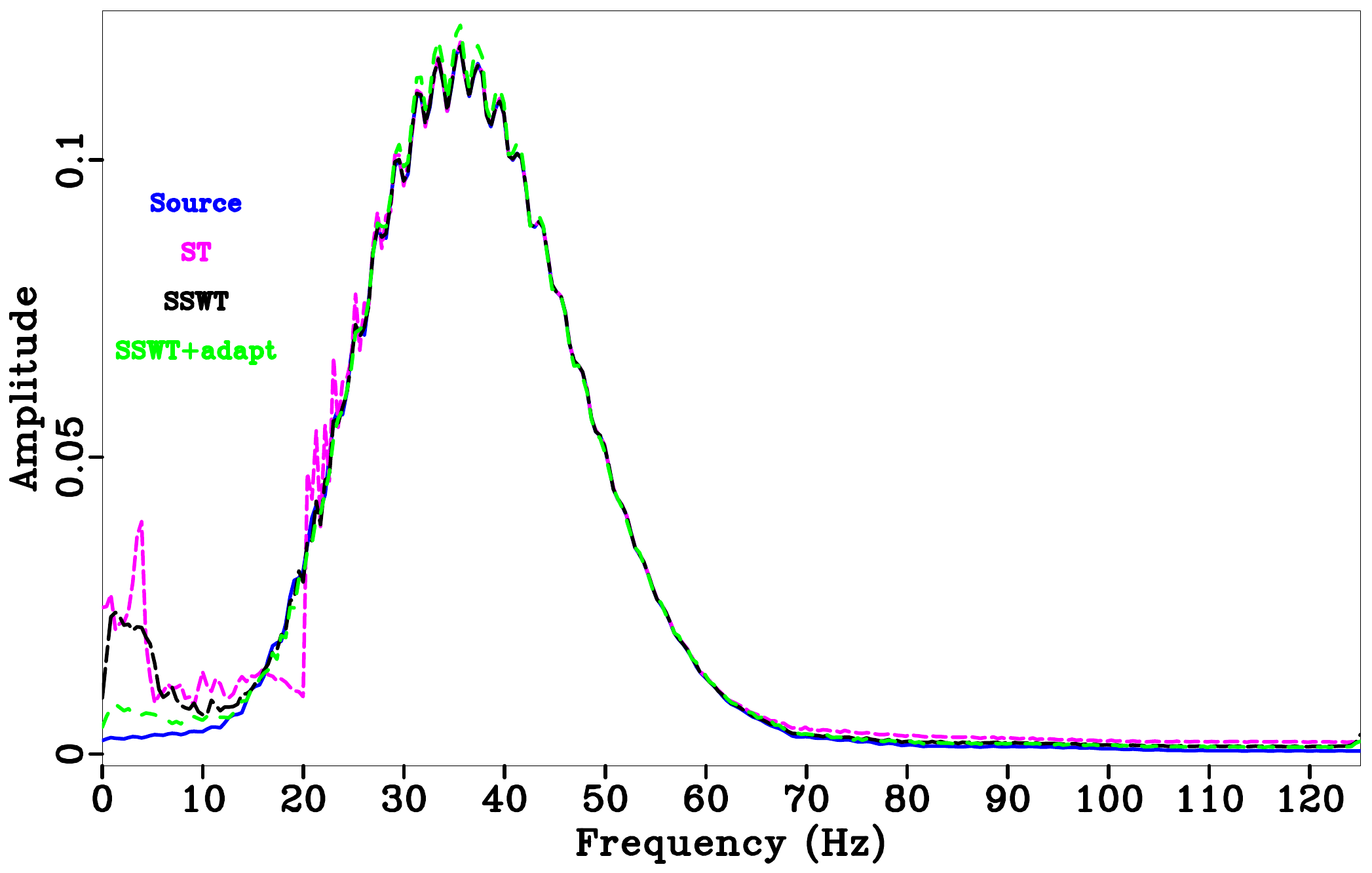}
            \caption{Comparison of averaged amplitude spectra obtained using different methods.}
            \label{fig:frequency_spectrum}
        \end{figure}

        Figure~\ref{fig:placeholder} displays the waveform of the 50th trace alongside its time-frequency (TF) representation computed via the synchrosqueezed wavelet transform (SSWT). The high-energy band distributed along the time axis, highlighted by the red rectangle, corresponds to the ground-roll energy, whereas the energy distributed primarily along the frequency axis, indicated by the arrows, represents the desired reflection events. For comparison, the synthesized data are processed using ground-roll suppression methods based on the seislet transform and the SSWT, with the corresponding separation results shown in Figures~\ref{fig:data_st} and~\ref{fig:data_st_ground}, respectively. The respective isolated ground-roll components obtained by these two methods are presented in Figures~\ref{fig:data_swt} and~\ref{fig:data_swt_ground}. As observed, when only localized TF filtering is applied, noticeable residual ground-roll energy remains in both scenarios. However, compared with the seislet-based method, the SSWT approach yields superior suppression performance owing to its significantly enhanced TF resolution. Additionally, severe energy leakage of the desired signals can be observed in the arrow-marked region of Figure~\ref{fig:data_swt}, which is primarily attributed to the inherent inverse reconstruction procedure of the seislet transform \cite{r18}. The separated data obtained using the proposed method, alongside the corresponding extracted ground-roll components, are illustrated in Figures~\ref{fig:data_adapt} and~\ref{fig:data_adapt_ground}, respectively. Following the application of the localized adaptive subtraction filter, the residual ground-roll energy in close proximity to the desired reflections is effectively attenuated. Figure~\ref{fig:frequency_spectrum} presents a comparison among the averaged amplitude spectra of the results obtained using the three evaluated methods and the ground-roll-free reference data. This comparison explicitly demonstrates that the proposed method exhibits superior suppression capability within the low-frequency ground-roll band, while simultaneously ensuring strict amplitude preservation of the desired signals compared to the conventional methods.

    \section{Field data application}
\begin{figure}
                \centering
                \begin{subfigure}{1\linewidth}
                    \centering
                    \includegraphics[width=1\linewidth]{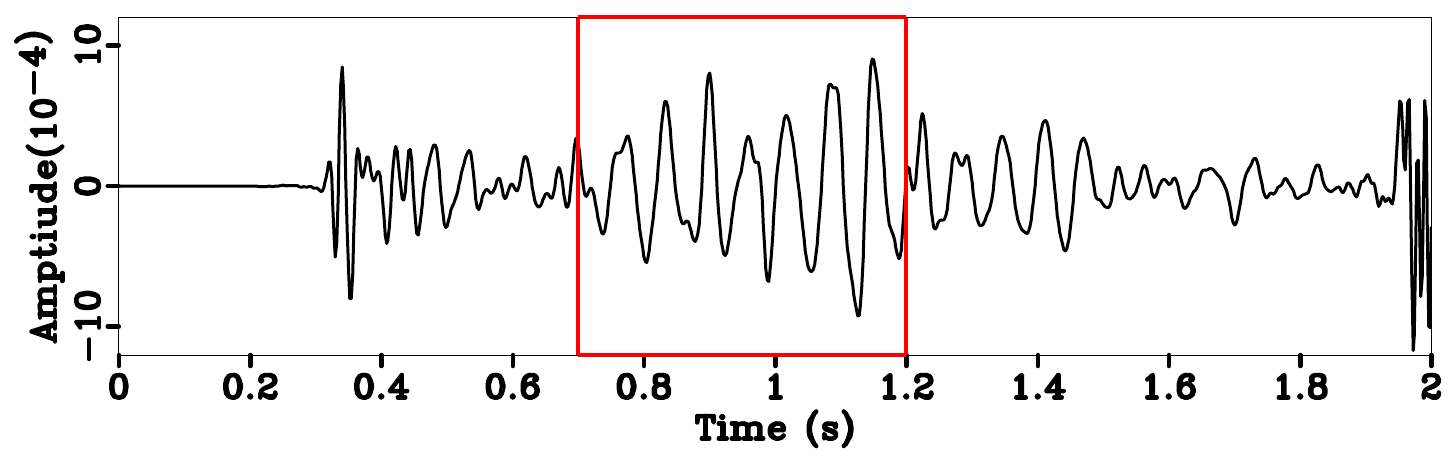}
                    \caption{}
                    \label{fig:field_trace550_signal}
                \end{subfigure}
                \begin{subfigure}{1\linewidth}
                    \centering
                    \includegraphics[width=1\linewidth]{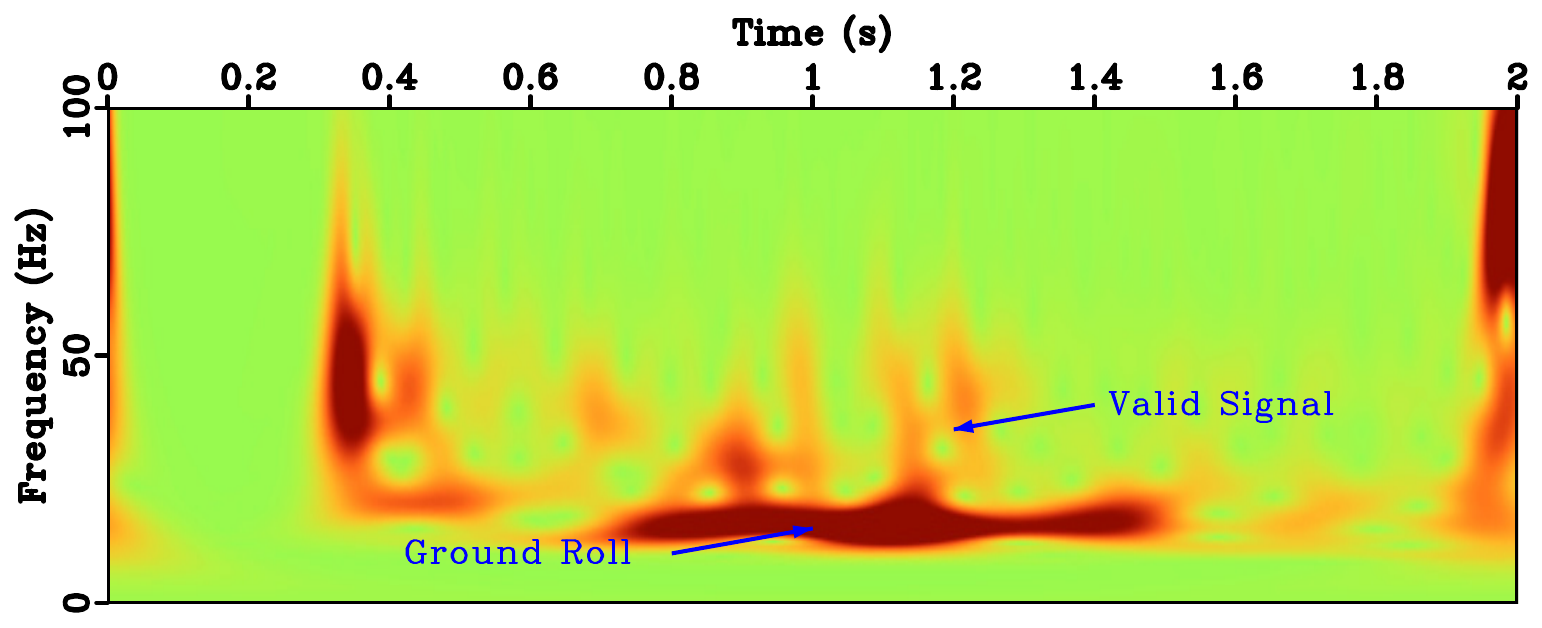}
                    \caption{}
                    \label{fig:field_trace550_tf_st}
                \end{subfigure}
                \begin{subfigure}{1\linewidth}
                    \centering
                    \includegraphics[width=1\linewidth]{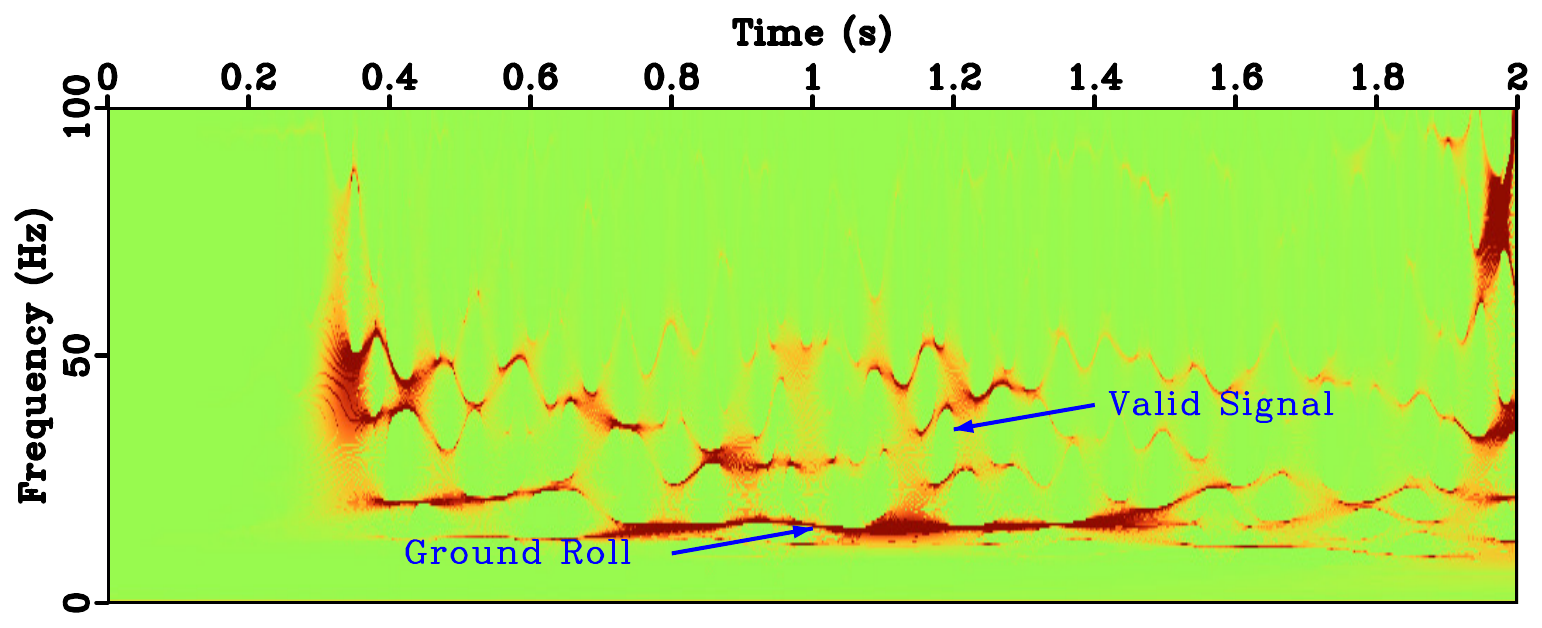}
                    \caption{}
                    \label{fig:field_trace550_tf_sswt}
                \end{subfigure}
                \caption{The 550th trace (a) and its time-frequency representations:(b) S-transform spectrum;(c) SSWT spectrum}
                \label{fig:Fig_field}
            \end{figure}

        To further evaluate the applicability and robustness of the proposed method in processing complex field seismic data, a representative shot gather from the Teapot Dome dataset, acquired in Wyoming, USA, is utilized. Specifically, the 214th shot gather is exclusively extracted for this analysis. The acquisition geometry comprises 17 receiver lines yielding a total of 966 traces. Each trace consists of 2049 temporal samples with a sampling rate of 2 ms, resulting in a total recording duration of approximately 4.1 s. The extracted shot gather is illustrated in Figure \ref{fig:Fig_field}. As observed, high-amplitude ground-roll energy severely contaminates the near-offset region and significantly overlaps with the desired reflection signals, thereby leading to a substantial degradation in the signal-to-noise ratio (SNR).

        The 550th trace is selected as a representative example for time-frequency (TF) analysis. The 1-D waveform and its corresponding TF representations computed via the conventional S-transform and the synchrosqueezed wavelet transform (SSWT) are displayed in Figures~\ref{fig:field_trace550_signal} and~\ref{fig:field_trace550_tf_st}, respectively. Compared with the S-transform, the SSWT spectrum exhibits a substantially higher TF resolution and a more highly concentrated energy distribution (Figure~\ref{fig:field_trace550_tf_sswt}). This superior TF concentration is highly advantageous for precisely identifying and isolating the ground-roll energy while strictly minimizing any potential damage to the desired reflections.

\begin{figure}[htbp]
    \centering
    \begin{subfigure}{0.45\linewidth}
      \centering
      \includegraphics[width=0.9\linewidth]{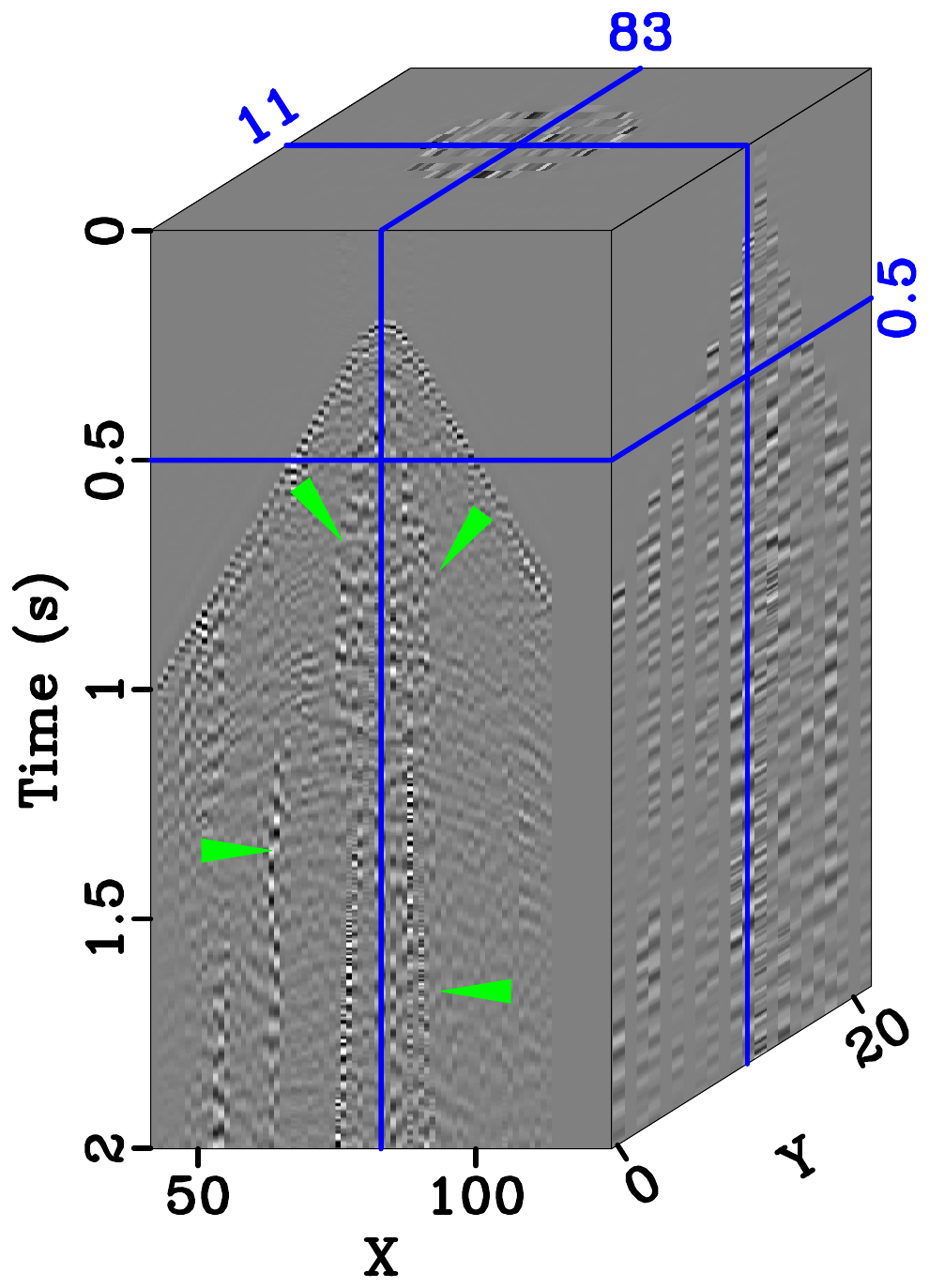}
      \caption{}
      \label{fig:ST-cube-box}
    \end{subfigure}
    \hfill
    \begin{subfigure}{0.45\linewidth}
      \centering
      \includegraphics[width=0.9\linewidth]{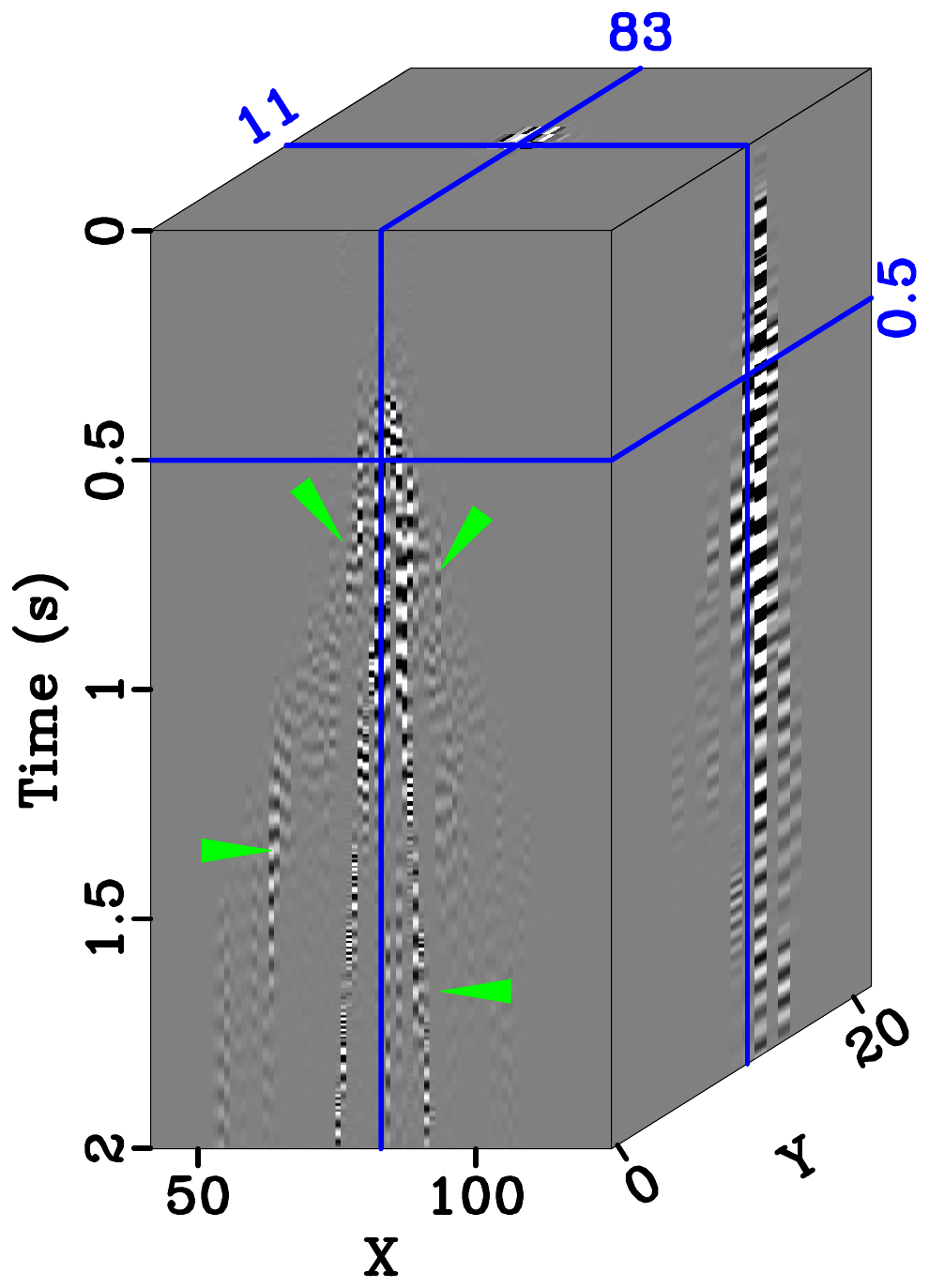}
      \caption{}
      \label{fig:ST_ground-cube-box}
    \end{subfigure}
    \vspace{0.3cm}    
    \begin{subfigure}{0.45\linewidth}
      \centering
      \includegraphics[width=0.9\linewidth]{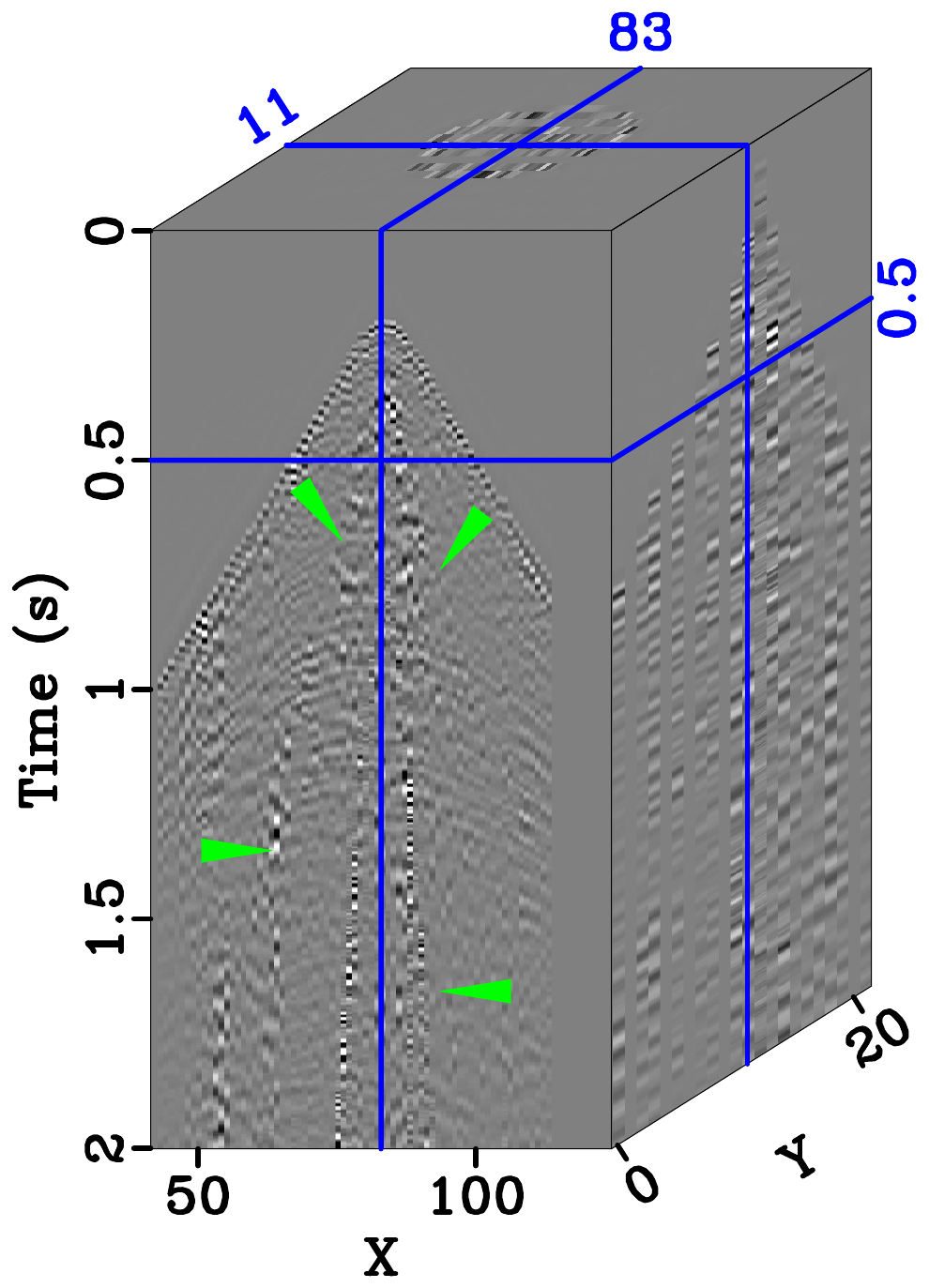}
      \caption{}
      \label{fig:SWT-cube-box}
    \end{subfigure}
    \hfill
    \begin{subfigure}{0.45\linewidth}
      \centering
      \includegraphics[width=0.9\linewidth]{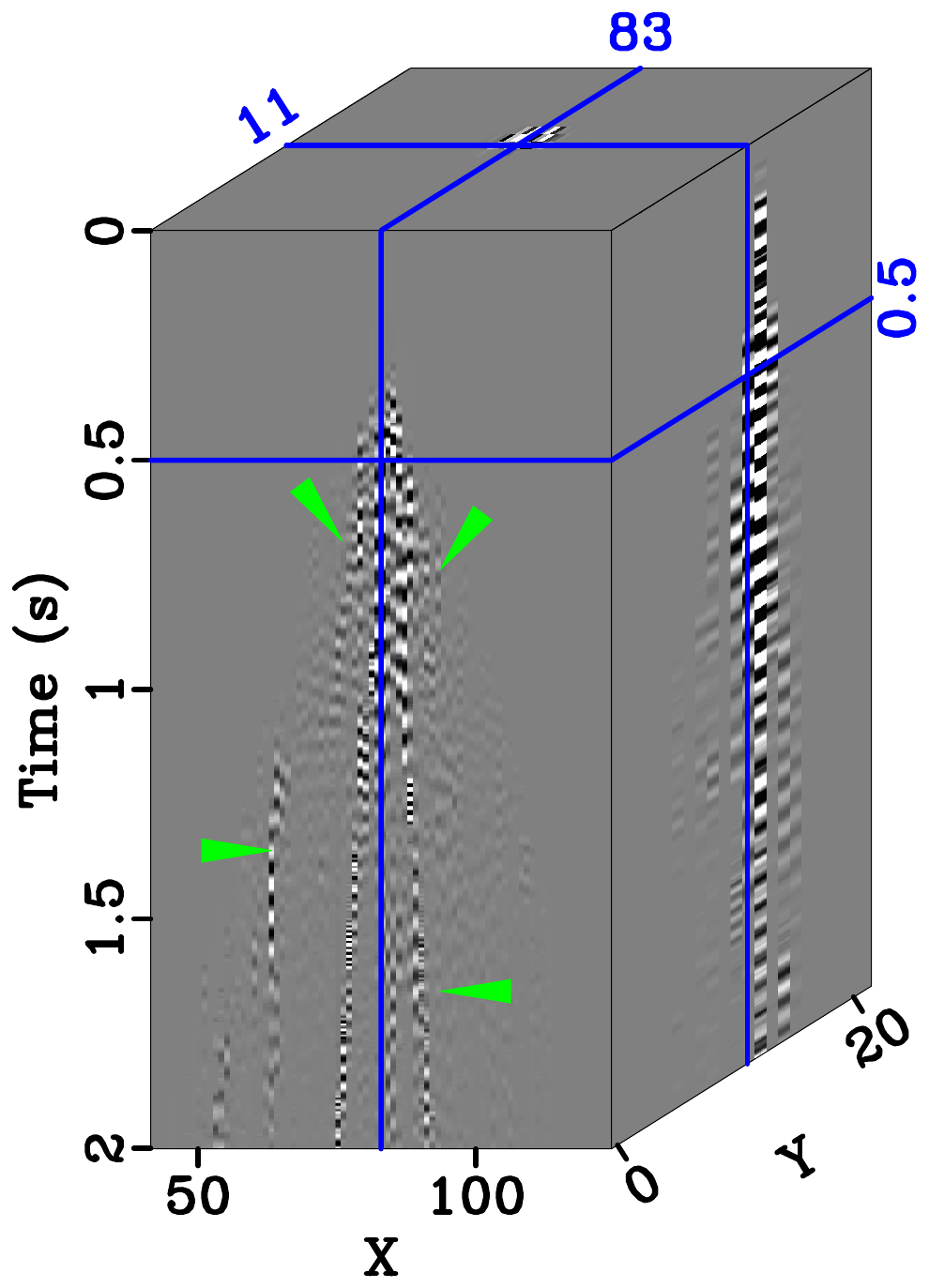}
      \caption{}
      \label{fig:SWT_ground-cube-box}
    \end{subfigure} 
    \vspace{0.3cm}
    \begin{subfigure}{0.45\linewidth}
      \centering
      \includegraphics[width=0.9\linewidth]{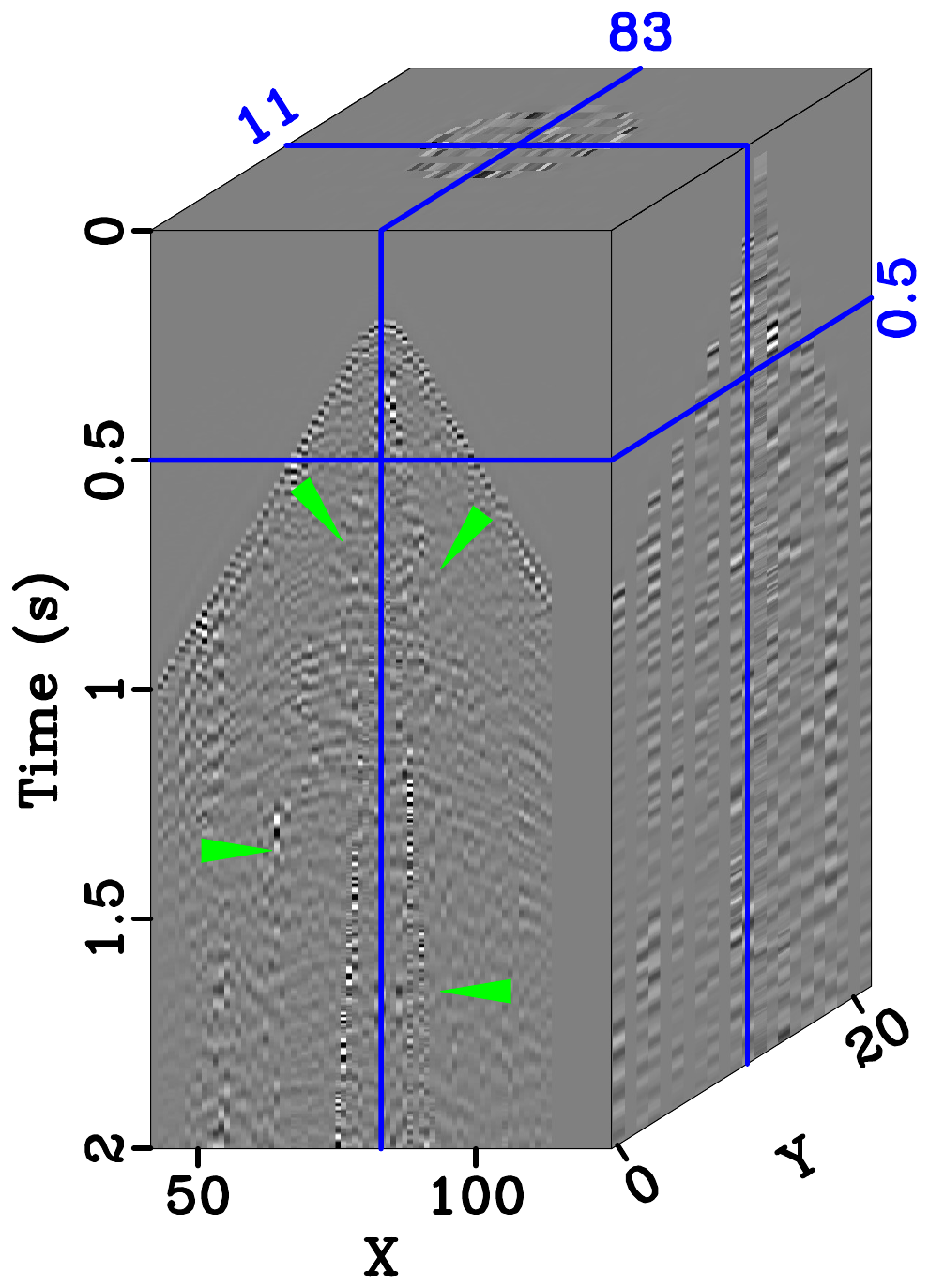}
      \caption{}
      \label{fig:SWT+adapt-cube-box}
    \end{subfigure}
    \hfill
    \begin{subfigure}{0.45\linewidth}
      \centering
      \includegraphics[width=0.9\linewidth]{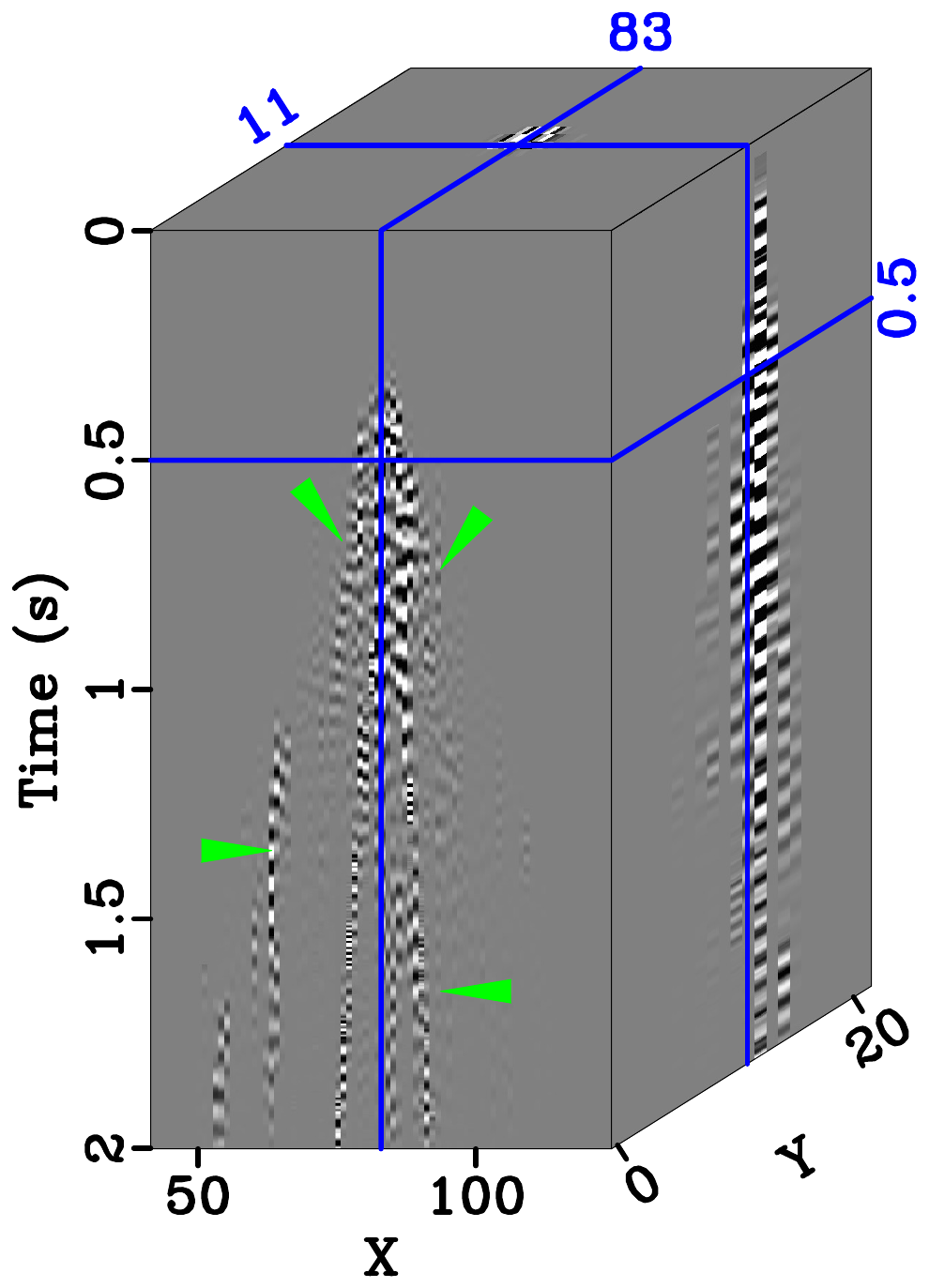}
      \caption{}
      \label{fig:SWT+adapt_ground-cube-box}
    \end{subfigure}
        \caption{Ground roll suppression results and extracted ground roll components obtained using different methods:(a), (d) S-transform; (b), (e) SSWT; (c), (f) proposed method.}
    \end{figure}

        Following the identification of the ground-roll-dominated regions within the time-frequency (TF) domain, localized TF filtering is executed utilizing both evaluated methods. The corresponding separated results are presented in Figures~\ref{fig:ST-cube-box} and~\ref{fig:SWT-cube-box}. As indicated by the highlighted regions, both the S-transform and the SSWT-based methods can partially attenuate the ground-roll energy. However, owing to the complex kinematics and high amplitudes of the ground roll in the near-offset region, noticeable residual interference remains after the application of conventional TF filtering, thereby leading to incomplete separation (Figures~\ref{fig:ST_ground-cube-box} and~\ref{fig:SWT_ground-cube-box}). Overall, the SSWT-based method significantly outperforms the S-transform-based approach due to its superior TF resolution.

        The separated data obtained via the proposed method are illustrated in Figure~\ref{fig:SWT+adapt-cube-box}. As explicitly observed, the ground-roll noise within the severely contaminated regions is effectively suppressed. Furthermore, the respective isolated ground-roll components obtained by the three evaluated methods are compared in Figures~\ref{fig:ST_ground-cube-box}, \ref{fig:SWT_ground-cube-box}, and~\ref{fig:SWT+adapt_ground-cube-box}. This comparison conclusively demonstrates that the proposed method yields the most superior performance in strictly preserving the desired reflections while optimally attenuating the ground-roll energy.

\section{Conclusions}
    Strong and dispersive ground roll in land seismic data often exhibits significant overlap with the desired reflection signals in the low-frequency band. This inherent interference makes it intrinsically difficult for conventional filtering techniques to achieve an optimal balance between noise attenuation and amplitude preservation. To rigorously address this challenge, a novel ground-roll attenuation framework integrating the synchrosqueezed wavelet transform (SSWT) and a localized adaptive subtraction filter is proposed in this study. The main conclusions can be summarized as follows:

    (1) The SSWT significantly enhances the separability between the ground roll and the desired reflection events within the time-frequency (TF) domain. Compared with the short-time Fourier transform, continuous wavelet transform, and S transform, the SSWT yields superior energy concentration and a substantially higher TF resolution, particularly for low-frequency components. This property enables the ground-roll energy to be explicitly represented as a continuous narrow-band structure, thereby providing a reliable mathematical basis for precise noise identification and isolation.

    (2) The localized TF thresholding strategy facilitates the construction of an estimated ground-roll model with strictly preserved dispersion characteristics and structural continuity. By exclusively retaining the TF regions dominated by ground-roll energy, severe energy leakage of the desired reflections into the predicted noise model is effectively prevented, thereby significantly improving the accuracy and reliability of the ground-roll modeling procedure.

    (3) The localized adaptive subtraction filter meticulously mitigates the kinematic and dynamic mismatches between the isolated ground-roll model and the actual acquired ground roll. By solving a regularized least-squares problem within localized spatiotemporal sliding windows, the adaptive filter precisely calibrates the amplitude and phase of the noise model in accordance with local variations. This yields optimally minimized residual interference while strictly preserving the amplitude fidelity and waveform continuity of the desired signals.

    (4) Extensive evaluations on both synthetic and field datasets conclusively demonstrate the efficacy and robustness of the proposed framework. Compared with conventional S-transform-based and CWT-based attenuation approaches, the proposed method exhibits a significantly enhanced suppression capability within the low-frequency band while ensuring strict amplitude preservation. Notably, even in the near-offset zones heavily contaminated by high-amplitude ground roll, the proposed method achieves substantial improvements in the signal-to-noise ratio (SNR) while maintaining the coherency of the underlying reflection events.

    In conclusion, the proposed framework provides an effective and robust solution for attenuating severe surface-wave noise while preserving the desired seismic signals in land data processing. Future research directions will primarily encompass: (1) automating the identification of ground-roll-dominated regions in the TF domain via machine learning or statistical feature extraction; (2) extending the SSWT to multi-trace or three-dimensional (3-D) seismic arrays to leverage spatiotemporal consistency; and (3) mitigating the inherently high computational intensity of the SSWT and multi-trace processing by exploiting GPU-based parallel acceleration, dimensionality reduction strategies, and fast approximate transforms for large-scale seismic applications.

\bibliography{reference}  
\bibliographystyle{IEEEtran} 
\end{document}